# Layer-by-layer growth of highly aligned $MoS_2$ nanoribbon arrays

Kaichi Yamamoto[1], Pablo Solís-Fernández,[2,*] Haiming Sun[3], Yanlin Gao[4,5], Yeri Lee[6], Yushan Tseng[7], Aika Uchida[2], Masahiro Hara[8], Sunmin Ryu[6], Yung-Chang Lin[3,9], Susumu Okada[4,5], Kazu Suenaga[3] & Hiroki Ago[1,2,10,*]

[1] *Interdisciplinary Graduate School of Engineering Science, Kyushu University, Fukuoka 816-8580, Japan*

[2] *Faculty of Engineering Science, Kyushu University, Fukuoka 816-8580, Japan*

[3] *SANKEN (The Institute of Scientific and Industrial Research), The University of Osaka, Osaka 567-0047, Japan.*

[4] *Department of Physics, Graduate School of Science and Technology, University of Tsukuba, Tsukuba 305-8577, Japan*

[5] *Tsukuba Institute for Advanced Research (TIAR), University of Tsukuba, Tsukuba 305-8565, Japan*

[6] *Department of Chemistry, Pohang University of Science and Technology (POSTECH), Gyeongbuk 37673, Republic of Korea*

[7] *Graduate School of Science and Technology, Kumamoto University, Kumamoto 860-8555, Japan*

[8] *Faculty of Advanced Science and Technology, Kumamoto University, Kumamoto 860-8555, Japan*

[9] *Research Institute of Core Technology for Materials Innovation, National Institute of Advanced Industrial Science and Technology (AIST), Tsukuba 305-8565, Japan*

[10] *Center for Semiconductors and Device Ecosystem (CSeDE), Kyushu University, Fukuoka 816-8580, Japan*

[*]Corresponding authors: Hiroki Ago (ago.hiroki.974@m.kyushu-u.ac.jp), Pablo Solís-Fernández (solisfernandez.pablo.kyudai@gmail.com)

## ABSTRACT

One-dimensional transition metal dichalcogenides (TMD) nanoribbons (NRs) offer a promising route to aggressive channel-width scaling in nanoscale transistors. However, controlling their layer number, a key determinant of device performance, has remained elusive. Here, we demonstrate a chemical vapor deposition (CVD) approach for layer-by-layer growth of $MoS_2$ NRs on a low-symmetry sapphire surface. This approach enables the oriented growth of single-crystalline bilayer NRs with a 2H stacking configuration, while maintaining widths below 20 nm. Increasing the $MoO_3$ precursor supply was found to be crucial for promoting the growth of the second and subsequent layers. By controlling the growth time and precursor supply, the layer number of $MoS_2$ NRs can be increased up to four layers. This method can also be extended to the growth of vertical $MoS_2/WS_2$ hetero-NRs. The average carrier mobility and current density of bilayer NR field-effect transistors (FETs) are two- and three-times higher than those of monolayer counterparts, respectively, and a maximum mobility of 86.7 $cm^2V^{-1}s^{-1}$ was attained with a bilayer NR-FET. These results highlight layer number as a key parameter for optimizing NR device performance. Our thickness-controlled growth strategy provides a general route towards TMD NR-based transistors with enhanced scalability and performance.

Transition metal dichalcogenides (TMDs), such as $MoS_2$ and $WSe_2$, have emerged as promising two-dimensional (2D) semiconductor materials for next-generation field-effect transistor (FET) architectures[1,2]. The transition from FinFETs to gate-all-around (GAA) FETs, and further to complementary FET (CFET) architectures, has been driven by the need for improved electrostatic channel control, reduced short-channel effects, and enhanced scalability. Owing to their atomically thin nature and dangling-bond-free surfaces, TMDs are expected to enable ultimate channel-thickness scaling in GAA and CFET architectures while maintaining high carrier mobility and strong gate controllability. Furthermore, channel-width scaling is equally required for additional transistor scaling. In this context, aligned TMD nanoribbons (NRs), which have one-dimensional (1D) structures, are particularly attractive and important.

Recently, monolayer TMD NRs with controlled orientation, chirality, and high crystallinity have been achieved by chemical vapor deposition (CVD) methods[3–8]. The orientation of CVD-grown TMD NRs reported in the literature has been driven by two main mechanisms: step- and lattice-guided growth. Surface step edges play a critical role in achieving oriented NR growth. On substrates, such as Au[3,4] and $Ga_2O_3$[5], step edges act as nucleation sites, with NRs growing along them to form a ribbon-like morphology. In the vapor-liquid-solid (VLS) method, the unidirectional movement of seed particles along step edges enables oriented NR growth[7]. In contrast, lattice-guided growth on low-symmetry sapphire has enabled the synthesis of oriented $MoS_2$ NRs with more precise edge control, greater versatility, and high density in a large area[8]. In addition, this lattice-oriented growth does not require controlled step/terrace structures. Compared with top-down approaches, such as lithography-based etching[9,10], the above CVD methods offer many key advantages, including well-controlled edge structures, the avoidance of fabrication-induced damages, and high productivity and throughput[11,12]. However, these bottom-up approaches have been predominantly limited to monolayer growth.

Bilayer TMD NRs are expected to offer several advantages over their monolayer counterparts, as demonstrated in bilayer CVD films. In terms of device performance, the bilayer structure offers higher carrier mobility and current density, as well as improved electrical contact[13–15]. The enhanced mechanical strength of the bilayer structure helps to maintain channel integrity in GAA transistors[16]. The synthesis of bilayer NRs has so far been demonstrated only by the VLS method[17]. However, the second layer does not fully cover the first layer, requiring additional etching step to obtain a well-defined bilayer NR. A method for the direct synthesis of NRs with controlled layer number has yet to be established, representing a critical challenge in TMD NR synthesis.

Here, we demonstrate the selective layer-by-layer growth of $MoS_2$ NRs on a-plane $(11\bar{2}0)$ sapphire. Our strategy enables the selective growth of additional layers on top of the grown monolayer, unlike the growth mode observed on c-plane sapphire, in which the monolayer continues to expand until full coverage of the substrate[18–20]. DFT calculations suggest that this behavior arises from the relatively weak interaction between a-plane sapphire and $MoS_2$. By leveraging lattice-guided monolayer NRs as templates, we successfully obtained dense arrays of bilayer $MoS_2$ NRs, while the layer number can be increased up to four by adjusting the growth conditions. Owing to their single crystallinity and well-controlled edges, our bilayer NRs exhibit notable device performance even at a narrow channel width down to 30 nm, surpassing that of monolayer NRs. Furthermore, this approach is not limited to $MoS_2$ NRs and can be extended to the fabrication of vertical $WS_2$/$MoS_2$ hetero-NRs. Our approach offers a fabrication route that enables control over the number of layers of aligned TMD NRs, significantly expanding the potential applications of TMD NRs.

## Results

### Synthesis and characterization of bilayer $MoS_2$ NRs

Bilayer $MoS_2$ NRs were grown on a-plane sapphire by switching the growth mode from in-plane to vertical layer-by-layer growth mode during the CVD process. This transition is triggered by an increase of the $MoO_3$ supply at elevated temperatures, as illustrated in Fig. 1a.

First, monolayer $MoS_2$ NRs were grown via substrate-lattice-guided epitaxy on the low-symmetry a-plane sapphire[8], resulting in aligned growth along the $[1\bar{1}00]$ direction of the sapphire (Fig. 1b). This yielded uniform monolayer NRs, as shown in the atomic force microscope (AFM) image in Fig. 1c. Subsequently, a second layer was selectively grown on the monolayer NRs in a layer-by-layer manner, with the increase of thickness controlled mainly by increasing the supply of $MoO_3$ precursor. Fig. 1d shows an AFM image of uniform bilayer $MoS_2$ NRs aligned on a-plane sapphire, demonstrating selective control of the layer number of the ribbons. To gain deeper insight into the growth mechanism of bilayer NRs, we performed automated segmentation of the AFM topography maps, extracting statistical information on the width, length, thickness, and orientation of the NRs (see Supplementary Fig. 1 for details). The result of the automated segmentation is displayed in Fig. 1e,f. The bilayer NRs exhibit average width and length of ~95 nm and ~1120 nm, respectively, which are comparable to those of the monolayer NRs despite the increase in the growth time. This reveals the preferential vertical growth mode over the in-plane growth mode of the NRs. Importantly, the bilayer NRs retain the high degree of alignment of the monolayer NRs along the $[1\bar{1}00]$ direction, as confirmed by orientations obtained from more than 240 NRs (Supplementary Fig. 2).

The increase in the layer number of the $MoS_2$ NRs was confirmed by photoluminescence (PL) and Raman spectroscopies. Fig. 1g,h show AFM images of a monolayer and a bilayer NR with similar widths and lengths, together with PL intensity mappings acquired from the same regions. The mappings reveal a decrease in PL intensity for the bilayer NR as compared with the monolayer NR, reflecting the transition from direct to indirect bandgap[21–23]. Supplementary Fig. 3 shows large-area PL intensity mappings along with the corresponding AFM images, corroborating the decrease in PL for the bilayer NRs. In the representative PL spectra shown in Fig. 1i, an additional peak at ~1.5 eV can be seen only in the bilayer NR, which originates from an indirect transition in multilayer $MoS_2$[13,22,23]. Raman spectroscopy also allows to determine the layer number of $MoS_2$ from the separation of the characteristic $A_{1g}$ and $E_{2g}$ peaks of $MoS_2$:, which increases with the layer number[24,25]. Fig. 1j shows mappings of this peak separation measured for the monolayer and bilayer NR arrays with peak separations of ~18 $cm^{-1}$ and ~22 $cm^{-1}$, respectively, which validate the layer number determined by AFM (see Supplementary Fig. 3 for the corresponding AFM image). In addition to bilayer NRs, a small fraction of trilayer NRs exist in the sample, as can be seen in the AFM image (Supplementary Fig. 3b), which has a peak separation of ~23.5 $cm^{-1}$ in the Raman mapping. Trilayer NRs exhibited a redshift of the indirect-transition PL peak as compared with the bilayer NR, reflecting the decrease in the indirect bandgap with the $MoS_2$ thickness (Supplementary Fig. 4)[21,22]. These results indicate the formation of aligned bilayer and trilayer $MoS_2$ NRs with a quality high enough to exhibit layer-dependent optical properties.

Cross-sectional scanning transmission electron microscopy (STEM) and transmission electron microscopy (TEM) analyses further confirmed the increase in the number of layers. Fig. 1k–m and Supplementary Fig. 5 show cross-sectional STEM and TEM images of monolayer, bilayer, and trilayer NRs sectioned perpendicular to their longitudinal axes. The observed sapphire surfaces correspond to the m-plane, regardless of the number of layers, proving that the NRs grow parallel to the $[1\bar{1}00]$ direction. It is worth noting that the influence of the sapphire steps on the alignment of the NRs is negligible under our growth condition, as evidenced by AFM image (Supplementary Fig. 6) that show a misalignment between the NRs and the steps having ~0.25 nm height. The STEM images shown in Figs. 1l and 1m also reveal the 2H stacking of the bilayer and trilayer NRs, as discussed below.

As already mentioned, the layer-by-layer mode allows growing NRs thicker than bilayer. AFM images in Fig. 1n show uniform single- to four-layer NRs, confirming that the height of each additional layer is ~0.6 nm, as expected for $MoS_2$. The determined layer numbers are further supported by Raman spectra, which indicate a gradual increase in separation of Raman $A_{1g}$ and $E_{2g}$ peaks from 18 $cm^{-1}$ for monolayer NRs to 24.5 $cm^{-1}$ for four-layer NRs (Fig. 1o).

To understand the mechanism underlying the increase in thickness, we investigated the time evolution of the second and additional layers during the CVD. Monolayer NRs were grown at the usual $MoO_3$ temperature of 600 °C, and then the second layer growth was initiated by elevating the $MoO_3$ temperature up to 640 °C. Fig. 2a–c displays a series of AFM images of $MoS_2$ NRs measured after different growth periods, allowing us to follow the progression of the second layer. After keeping the temperature of $MoO_3$ at 640 °C for 10 min, only a fraction of the NRs was uniform monolayer (13%) or bilayer (21%), whereas the majority exhibited a mixed morphology consisting of both monolayer and bilayer regions. The fractions of the uniform NRs with each layer number are summarized in Supplementary Fig. 7.

It is noted that at the initial stage, the second layer nucleates selectively on top of the existing monolayer NRs. In these partially thickened NRs, bilayer domains cover substantial fractions of the ribbon areas, ranging from isolated bilayer domains nucleated on a monolayer ribbon to ribbons that are almost entirely bilayer. This suggests a progressive expansion and coalescence of bilayer domains during growth. The trend is confirmed by the AFM image taken after 20 min, in which the fraction of NRs that are fully covered bilayer ribbons has significantly increased (Fig. 2b). At this stage, small trilayer domains begin to appear on a few bilayer NRs, indicating that nucleation of the third layer starts only when the second-layer coverage is almost complete. Further reaction with $MoO_3$ for 45 min converts almost all of the monolayer NRs to bilayer and a few trilayer ribbons (Fig. 2c). These observations signify that the number of NR layers increases in a layer-by-layer manner. Fig. 2d summarizes the evolution of fraction of the monolayer, bilayer, and trilayer NRs. The fraction of NRs classified as bilayer (*i.e.*, ribbons in which the bilayer domain occupies most of the ribbon area) increases with growth time and reaches a high abundance of 92 % at 45 min, realizing the production of a predominantly bilayer NR array.

Fig. 2e compares the distributions of NR widths and lengths at each growth time, revealing that NR sizes do not change significantly over extended growth times. This result indicates that vertical growth of NRs is preferred over in-plane growth, demonstrating that our approach enables layer-number control while preserving the narrow width of the NRs. The bilayer NRs can have widths down to ~20 nm, as measured by AFM (Supplementary Fig. 8). Because AFM measurements overestimate the NR width due to tip convolution, as discussed in Supplementary Fig. 9, the narrowest bilayer NRs are expected to have actual widths below 20 nm.

This observed layer-by-layer growth behavior clearly differs from that of bilayer TMDs grown on the sapphire substrates having a step-terrace structure with controlled step height in previous studies[14,15]. In our approach, raising the $MoO_3$ temperature promotes the growth of the additional layers due to an increase of the Mo flux. As previously demonstrated, the formation of bilayer and few-layer $MoS_2$ requires a higher Mo flux than that needed for monolayer growth[13,26]. Moreover, a high metal/chalcogenide ratio has been reported to promote vertical growth rather than in-plane growth[27].

It is interesting that the layer-by-layer growth proceeds on a-plane sapphire without covering the whole sapphire surface with a monolayer $MoS_2$ film. We performed DFT calculation to get further insight into the growth mechanism underlying this limited vertical growth. Adsorption energies of small $MoS_2$ flakes deposited on the surfaces of $MoS_2$ and on a-plane sapphire were calculated, as illustrated in Fig. 2f and Supplementary Fig. 10. The calculations reveal that the adsorption energy on $MoS_2$ is -1.53 eV, which is lower than that on a-plane sapphire (-1.04 eV). This indicates that once a $MoS_2$ NR forms, further nucleation on top of the existing $MoS_2$ is energetically more favorable than on the a-plane sapphire. Thus, we speculate that the selective growth observed here is attributed to the relatively weak interaction between a-plane sapphire surface and $MoS_2$[28]. In contrast, on c-plane sapphire, nucleation of the second layer is suppressed until the first layer fully covers the substrate surface[18–20].

As discussed above, the $MoO_3$ temperature is the key factor driving the increase in NR thickness, rather than the growth duration. Hence, even when the growth time was extended from 10 min to 20 min while keeping the $MoO_3$ temperature at 600 °C, the NRs remained monolayer (Supplementary Fig. 11). To study the impact of $MoO_3$ temperature, it was varied from 600 °C to 660 °C while keeping the total growth time fixed at 45 min (the exception was the 600 °C sample, for which the total growth time was 10 min to obtain monolayer NRs (Fig. 3a)). When the $MoO_3$ was heated to 620 °C, the second layer only partially covered the monolayer NRs (Fig. 3b). Increasing the temperature to 640 °C resulted in nearly complete bilayer coverage, with 77% of the NRs being bilayer (Fig. 3c). At these temperatures a small fraction of trilayer NRs was also present. A further increase to 660 °C resulted in an increase in the ratio of trilayer NRs up to 5% (Fig. 3d). The fractions of NRs for each layer number under different $MoO_3$ temperatures are summarized in the corresponding pie charts (insets of Fig. 3).

These observations suggest that increasing the $MoO_3$ supply by increasing the precursor temperature promotes the successive layer growth and enhances the growth kinetics of each additional layer, as the layer number increases despite the identical growth duration. Consequently, a $MoO_3$ temperature around 640 °C is proved to be optimal for the growth of uniform bilayer NRs, while higher temperatures favor the growth of thicker NRs owing to the layer-by-layer mechanism described above. Motivated by this trend, we further extended the growth time at 660 °C from 45 min to 75 min to obtain thicker NRs. As shown in Fig. 3e, a fourth layer started to form on top of the trilayer. For the area shown in Supplementary Fig. 12, 27.8% of the NRs are classified as monolayer, 43.9% as bilayer, 26.0% as trilayer, and 2.3% as four-layer NRs. Despite the formation of trilayer and four-layer NRs at the higher temperatures, it is worth mentioning that monolayer NRs are still observed and even increase their proportion when the temperature is increased from 640 °C to 660 °C (see pie charts in Fig. 3, and Supplementary Fig. 12 for the distribution of widths and lengths of the NRs). These monolayer NRs are narrower and shorter than those grown at 600 °C (see Fig. 1e), suggesting that the NR nucleation at high temperatures continued even at the later stages of the growth so that they do not have sufficient growth time to develop into thicker NRs.

**Structural characterizations and heterostructure**

The atomic structures of bilayer and trilayer NRs, their edges, and their stacking configurations were characterized by STEM. Fig. 4a displays a top-view STEM image of a 35-nm-wide bilayer NR. A high-magnification image (Fig. 4b) confirms that its longitudinal axis is aligned parallel to the zigzag direction of the $MoS_2$ lattice. In most cases, the edges of the first and second layers coincide, as shown in Fig. 4b, indicating that the second layer extends to the edge of the first layer. These high-resolution STEM images reveal the absence of structural defects, indicating a high crystallinity of the NRs. Furthermore, no grain boundaries are observed in the second layer, even though the second layer can simultaneously nucleate at multiple locations of a given ribbon. Together with the single-crystalline nature of the bottom monolayer NRs, these observations confirm that the bilayer NRs are single-crystalline. Grain boundaries have previously been reported to form during the merging of regions with different stacking orders[14,15,29]. However, the NRs here exhibit only a 2H stacking order, resulting in single crystallinity, as will be discussed below.

Fig. 4c shows a STEM image of an edge of a bilayer NR, in a region where the edges of the first and second layers do not coincide, allowing the study of the different edges separately. The STEM analysis reveals that the edge of the first layer is Mo-zigzag (ZZ-Mo), whereas the straight edge of the second layer is S-zigzag (ZZ-S). In some regions, the S-terminated edge of the second layer is reconstructed into a ZZ-Mo edge, as shown in Supplementary Fig. 13. These observations are consistent with theoretical calculations predicting that the formation energy of the ZZ-Mo edge decreases under Mo-rich growth conditions[30,31]. The contrast of the image increases with the number of layers, and the stacking configuration in the bilayer region is identified as 2H (see Supplementary Fig. 14 for more detailed analysis). The same 2H stacking configuration was also observed in the trilayer region (Supplementary Fig. 15). These experimental images are reproduced by simulations based on the 2H $MoS_2$ structure, as shown in Fig. 4c. Corresponding atomic-resolution elemental maps measured by electron energy-loss spectroscopy (EELS) further confirm that the monolayer and bilayer regions can be distinguished by their contrast differences, and that Mo and S are present in each atomic column in the bilayer (Fig. 4d,e). A low-magnification STEM image shown in Supplementary Fig. 16 shows monolayer and bilayer NRs, with an isolated trilayer grain observed on top of the bilayer NR, further supporting the conclusion that the NRs grow in a layer-by-layer manner.

The stacking configuration of bilayer NRs was further confirmed by second-harmonic generation (SHG) measurements. The SHG intensity from bilayer NRs was found to be significantly weaker than that from monolayer NRs, despite their similar widths (Fig. 4f). This suppressed intensity in bilayer NRs can be attributed to the centrosymmetric 2H stacking configuration, in contrast to the broken centrosymmetry of the monolayer, which exhibits strong SHG signals[32–35]. As shown in Supplementary Fig. 17, the SHG of monolayer NRs showed an increase in intensity with width, since the NRs are typically smaller than the laser spot size (~500 nm). In contrast, the bilayer NRs exhibited no clear dependence of SHG intensity on their widths, indicating that the suppressed SHG intensity does not originate from the limited size of the NRs, but rather from their 2H stacking configurations.

Our sequential layer-by-layer growth strategy demonstrated here can be readily extended to the fabrication of vertical NR heterostructures. Such heterostructures have previously been synthesized by epitaxial growth[36], defect engineering[29,37], and chalcogenization of deposited metal[38]. Here, the $MoO_3$ precursor was replaced with $WO_3$ after forming bilayer $MoS_2$ NRs, enabling the selective growth of $WS_2$ on top of $MoS_2$ NRs, resulting in vertical $MoS_2/WS_2$ hetero-NRs (Fig. 5a). Fig. 5b shows an AFM image of a heterostructure NR, in which $WS_2$ covers a part of a bilayer $MoS_2$ NR, facilitating the interpretation of the subsequent structural and spectroscopic analysis. The height of the heterostructure region corresponds to that expected for a trilayer, consistent with the addition of a single $WS_2$ layer on top of the bilayer $MoS_2$. The presence of $WS_2$ was confirmed by Raman measurements performed on the identical NR. The intensity mappings of the $MoS_2$ $A_{1g}$ and $WS_2$ 2LA(M) Raman modes show that $MoS_2$ is present throughout the NR, while $WS_2$ is localized to the thicker region (Fig. 5c,d). The spectrum of the $MoS_2$-only region exhibits the characteristic $E_{2g}$ and $A_{1g}$ peaks of $MoS_2$, whereas the spectrum from the heterostructure region displays additional peaks originating in $WS_2$ (Fig. 5e)[39].

Cross-sectional annular dark field STEM (ADF-STEM) images further revealed a layer with distinct contrast stacked on top of the bilayer $MoS_2$, confirming the formation of a vertical heterostructure (Fig. 5f). Elemental analysis by EELS, shown in Fig. 5g, confirms that the top layer is $WS_2$, with W and Mo confined to their corresponding layers without detectable elemental intermixing across the interface. This proves that the sequential layer-by-layer growth strategy is not limited to homostructures but is also applicable to the controlled fabrication of 1D vertical TMD heterostructures.

**Electrical properties of bilayer NRs**

The carrier transport properties of bilayer and monolayer NRs were evaluated by fabricating back-gated FET devices. To directly compare the device performance, we patterned channels within a single 70-nm-wide NR containing both bilayer and monolayer regions, as shown in Fig. 6a inset and Supplementary Fig. 18. This allows us to study the influence of layer thickness while maintaining an identical ribbon width. Since the NR length is typically around ~1 μm, the channel length for these devices was set to 340 nm. Fig. 6a exhibits transfer curves of the monolayer and bilayer channels with Au/Ni electrodes. Both showed n-type carrier transport and similar voltage thresholds, but the bilayer channel showed a much higher on-current. The electron mobilities of the monolayer and bilayer FETs are 3.5 $cm^2V^{-1}s^{-1}$ and 17.6 $cm^2V^{-1}s^{-1}$, respectively, indicating an improved device performance for the bilayer.

While these short-channel devices allow a direct comparison under identical channel widths, their performance is expected to be strongly influenced by contact resistance. Therefore, to better evaluate the intrinsic transport properties of the bilayer NRs and compare the device performance between monolayer and bilayer NRs, we fabricated back-gated FETs with a channel length of 900 nm. To minimize the fabrication-induced variability, monolayer and bilayer NRs were synthesized on the same sapphire substrate by tuning the growth condition and transferred simultaneously on the same $SiO_2$ substrate prior to FET fabrication. Fig. 6b shows the transfer curves of the FET fabricated using a 60-nm-wide bilayer channel, exhibiting mobility and current density of 71.3 $cm^2V^{-1}s^{-1}$ and 63.1 μA/μm, respectively. The output characteristics of this device showed linear $I_D$–$V_D$ curves (Fig. 6c), indicating a good ohmic contact between the NR and the electrodes. We note that this performance is comparable to that reported for wide-channel FETs despite the present channel width being below 100 nm[40], underscoring the high quality of the bilayer $MoS_2$ NRs.

Fig. 6d,e compare the mobility and current density of the monolayer and bilayer NR FETs. The bilayer devices exhibit an average mobility of 43.2 $cm^2V^{-1}s^{-1}$, approximately two times higher than that of the monolayer devices (22.1 $cm^2V^{-1}s^{-1}$). Likewise, the average current density at $V_G$ = 1 V is 30.1 μA/μm for the bilayer devices, nearly three times higher than that of the monolayer devices (10.3 μA/μm). These results demonstrate the clear advantage of bilayer NRs over their monolayer counterparts for high-performance devices. We speculate that the remarkable performance originates from the absence of grain boundaries, as bilayer NRs adopt only the 2H stacking configuration, and reduced effect from the underlying $SiO_2$ substrate. A previous study has reported degraded device performance at 2H/3R grain boundaries[14]. On the other hand, our bottom-up-grown bilayer NRs are free from the rough

edges typically introduced by etching processes in top-down methods. These factors may contribute to the high transport properties of the bilayer NRs.

We also fabricated back-gated bilayer $MoS_2$ NR FETs using Sb as a contact metal instead of the Ni to reduce the contact resistance[41]. The bilayer NR FETs with a channel length of 900 nm showed enhanced performance with mobility of 86.7 $cm^2V^{-1}s^{-1}$ and current density of 51.8 μA/μm, as shown in Supplementary Fig. 19, suggesting that further device optimization could lead to even higher performance.

Fig. 6f compares our results with previously reported data for the devices based on both top-down and CVD-grown NRs[4,5,10,17,42–44]. Direct comparison is complicated by differences in device geometry and measurement conditions, such as substantially shorter channel lengths, different dielectric substrates, or different drain voltages. Thus, we benchmark the field-effect mobility as a function of channel width to enable a broad comparison across the literature (see also Supplementary Table 1). Our monolayer NRs exhibit high field-effect mobility comparable to that of the best previously reported CVD-grown monolayer NRs. We note that the bilayer NRs exhibit field-effect mobilities comparable to, or even exceeding, those of the highest-performing TMD NR FETs reported to date. These results infer that the combination of the bilayer structure and the etching-free bottom-up growth approach is highly beneficial for achieving high device performance, highlighting the potential of bilayer $MoS_2$ NRs for future transistor scaling.

## Discussion

We demonstrated that controlling the Mo supply enables the successful growth of aligned bilayer and few-layer NRs on a-plane sapphire along the $[1\bar{1}00]$ direction. SHG measurements and STEM analysis clarified that the ribbons have a 2H stacking configuration, resulting in single-crystalline bilayer NRs with well-defined zigzag edges. Combined with the DFT results, systematic investigations of growth time and $MoO_3$ temperature revealed that NRs grow via a layer-by-layer mode while maintaining their narrow widths. This approach can be extended to fabricate vertical hetero-NRs by varying the precursor during growth. Furthermore, the transport measurements proved that bilayer NRs outperform their monolayer counterparts, highlighting the advantage of the bilayer structure for further device scaling.

## Methods

### CVD growth of bilayer $MoS_2$ NRs

$MoS_2$ NRs were synthesized using a three-zone furnace with a 26-mm inner-diameter quartz tube at ambient pressure, using Ar as the carrier gas at a flow rate of 300 sccm (standard cubic centimeters per minute). Sulfur (Mitsuwa Chemical Co., 99.999%, 120 mg) and $MoO_3$ (Wako Pure Chemical Industries, 99.9%, 14 mg) powder precursors were placed upstream and heated at 210 °C and 600 °C, respectively, after which the $MoO_3$ temperature was increased up to 640 °C. Sapphire a-plane $(11\bar{2}0)$ substrates (Kyocera Co.) with a small miscut angle of $<0.3°$ were cleaned by sonication in acetone and isopropanol, followed by drying with $N_2$ blow before CVD growth. The sapphire was placed downstream and heated at 1110 °C during the growth.

For the growth of $WS_2/MoS_2$ vertical heterostructure, sulfur and $WO_3$ (Kanto Chemical Co., 99.0%, 15 mg) powder were placed upstream and heated at 175 °C and 1120 °C, respectively. The sapphire substrate covered with $MoS_2$ NRs was placed downstream and heated at 950 °C. Growth was carried out under an Ar flow of 600 sccm for 10 min, followed by natural cooling of the furnace.

### Transfer of $MoS_2$ NRs

The bilayer $MoS_2$ NRs were transferred to a $SiO_2$ (300 nm)/Si substrate by the ultraviolet (UV) tape-assisted transfer method reported previously[45]. After the UV tape was attached to the as-grown NRs, it was illuminated with UV light, making the adhesive layer solid and significantly weakening the adhesion force[45]. The UV tape with the NRs was detached from the sapphire substrate in a 1 mol/L KOH solution, followed by washing in ultrapure water. Then, the UV tape/$MoS_2$ NRs stack was attached to a target $SiO_2$/Si substrate, followed by baking. Finally, the UV tape was slowly peeled off to leave the NRs on the $SiO_2$/Si substrate.

**Characterizations**

Raman and PL measurements were conducted under ambient conditions using a Nanofinder 30 spectrometer (Tokyo Instruments) with a 532-nm laser excitation and 0.1~0.3 mW. SHG measurements were performed using a home-built micro-SHG spectroscopy setup integrated into a microscope, as described previously[46], using an 800 nm Ti:sapphire pulsed laser (effective focal spot size: ~500 nm) at an average power below 0.2 mW. SEM (HITACHI S-4800) and AFM (Bruker Nanoscope Ⅴ) were used for characterization of NRs and devices. The cross-sectional specimens for STEM/TEM measurements were prepared with a focused ion beam (FIB) system (HITACHI NX-2000). The structural observations were carried out using JEOL JEM-ARM200F with a Cs corrector operated at 200 kV, and the elemental analysis was performed with Gatan Continuum. For top-view STEM measurement, $MoS_2$ NRs were transferred onto a TEM grid (Quantifoil) covered with graphene. Atomic resolution, top-view STEM images were conducted on TripleC#3 (JEOL, ARM200F-based ultra-high vacuum microscope) equipped with delta corrector and cold field emission gun operating at 60 kV. The convergence semi-angle of the probe and internal collection semi-angle of ADF detector were 37 and 76 mrad, respectively. ADF-STEM image simulations were carried out using the multislice method as implemented in the MacTempas, with simulation parameters set close to the experimental conditions. The corresponding EELS mappings were acquired using a Gatan system combining DigiScan 3 and GIF Continuum plus STELA hybrid-pixel camera, optimized for low-voltage operation. All STEM/TEM images and EELS were acquired at room temperature.

**Device fabrication and measurement**

Back-gated FETs with bilayer $MoS_2$ NRs were fabricated on a $SiO_2$ (300 nm)/Si substrate. Before spin-coating the resist, the transferred NRs were heated at 150 °C in a glove box filled with $N_2$ gas. Electron-beam lithography (ELIONIX Inc. ELS-S50) was used to form source and drain electrode patterns using a double-layer resist of PMMA 495K and PMMA 950K (both 6% by weight in anisole, spin-coated sequentially at 4000 rpm), followed by electron-beam evaporation of Au/Ni (30 nm/2 nm) and Au/Sb (30 nm/20 nm) electrodes. Device measurements were performed in vacuum (~$10^4$ Pa) and at room temperature using a semiconductor analyzer (Keysight B1500A). The field-effect mobility was extracted from the linear regime of the transfer characteristics using the following equation:

$$\mu_{\mathrm{FE}} = g_{\mathrm{m}} \frac{L}{W C_{\mathrm{OX}} V_{\mathrm{D}}} \text{ with } g_{\mathrm{m}} = \frac{\partial I_{\mathrm{D}}}{\partial V_{\mathrm{G}}}$$

where $L$ and $W$ are the channel length and width, respectively, $C_{\mathrm{ox}}$ is the gate oxide capacitance per unit area, $V_{\mathrm{D}}$ is the source-drain voltage, and $g_{\mathrm{m}}$ is the transconductance extracted from transfer curves

## DFT calculations

All calculations were performed based on the density functional theory[47,48] combined with the effective screening medium method[49]. The exchange-correlation potential energy of interacting electrons was treated using vdW-DF2 to describe weak dispersive interactions in van der Waals structures[50,51]. Ultrasoft pseudopotentials were used to describe the interaction between valence electrons and ions[52]. Valence wavefunction and deficit charge density were expanded by the plane-wave basis sets with cutoff energies of 25 and 225 Ry, respectively. Because of structural asymmetry perpendicular to the $\alpha$-$Al_2O_3$ and $MoS_2$ surfaces, we used the effective screening medium method to exclude unphysical electrostatic interactions with the imaging cells by imposing an open boundary condition for electrostatic potential at the cell boundaries.

## Data availability

The data that support the findings of this study are available from the corresponding author upon reasonable request.

**Acknowledgements**

This work was supported by JSPS Grant-in-Aid for Scientific Research on Innovative Areas "Science of 2.5 Dimensional Materials: Paradigm Shift of Materials Science Toward Future Social Innovation" (KAKENHI grant numbers JP21H05232, JP21H05233, 21H05235), JSPS KAKENHI grant numbers JP26K22696, 26H00004, 25K01748, JP24H00407, JP25K08435, JST-Research and Development Program for Next-generation Edge AI Semiconductors Grant Number JPMJES2531, JST-CREST grant number JPMJCR20B1, and ERC Synergy grant (MORE-TEM 951215). K.Y. acknowledges support from JST K-SPRING program (grant number JPMJSP2136). S.R. acknowledges support from the National Research Foundation of Korea (NRF) (RS-2024-00336324).

**Author contributions**

K.Y., P.S.-F., and H.A. conceived the idea and designed the experiments. K.Y. and A.U. performed the synthesis of the NRs. K.Y., and P.S.-F. transferred the as-grown NRs to $SiO_2$ and to graphene/TEM grids. H.S., Y.-C.L., and K.S. performed the top-view STEM and EELS measurements. Y.L. and S.R. measured SHG. S.O. and Y.G. performed DFT calculations. K.Y., Y.T., and M.H. performed EB patterning and the FET measurements and analyzed the results. P.S.-F. and H.A. supervised this project. K.Y., P.S.-F., and H.A. wrote the manuscript, with all authors providing input and comments on the manuscript.

**Competing interests**

The authors declare no competing interests.

**Additional information**

**Supplementary information** The online version contains supplementary material available at https://doi.org/10.1038/.

**Correspondence and requests for materials** should be addressed to Hiroki Ago.

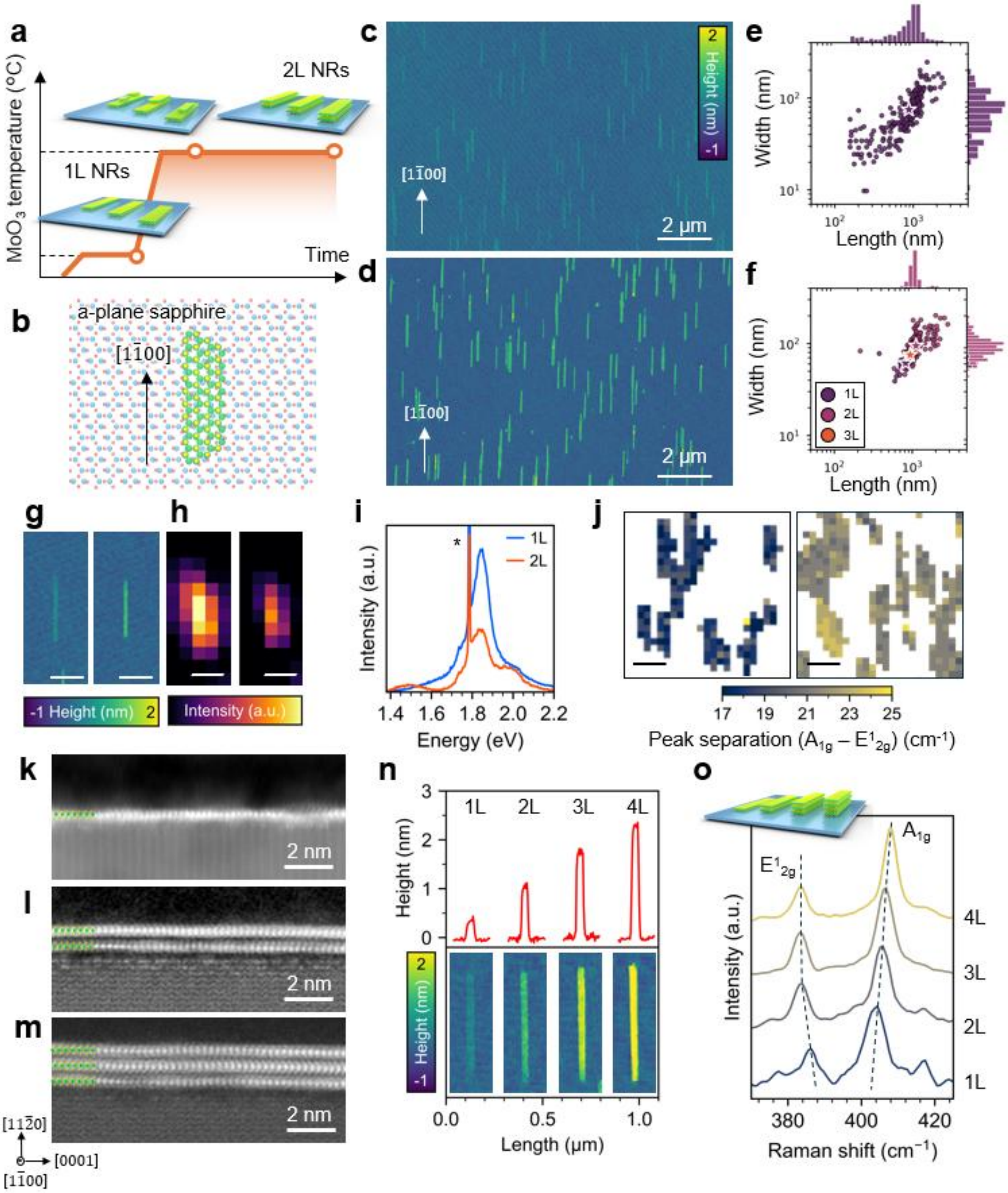

**Fig. 1. Thickness-controlled growth of aligned $MoS_2$ NRs on a-plane sapphire. a** Temperature profile of $MoO_3$ during the CVD growth used for selective growth of monolayer (1L) and bilayer (2L) NRs. **b** Atomic model of a $MoS_2$ NR aligned on a-plane sapphire. AFM images of monolayer (**c**) and bilayer (**d**) $MoS_2$ NR arrays measured on the sapphire. Distributions of length and width of monolayer (**e**) and bilayer (**f**) NRs, determined by segmentation of AFM images. Stars indicate the average values for each layer number. AFM images (**g**) and PL intensity mappings (**h**) of monolayer (left) and bilayer (right) NRs. Scale bars are 500 nm. **i** PL spectra of as-grown monolayer (blue) and bilayer (red) NRs. The asterisk

indicates the peak from the sapphire substrate. **j** Raman peak separation mappings of as-grown monolayer (left) and bilayer (right) NRs, with a trilayer NR. Scale bars are 1 μm. **k–m** Cross-sectional STEM images of monolayer (top), bilayer (middle), and trilayer (bottom) NRs on a-plane sapphire. **n** AFM images and line profiles of isolated NRs with thickness ranging from one to four layers. **o** Raman spectra of one- to four-layer $MoS_2$ NRs with a schematic on top.

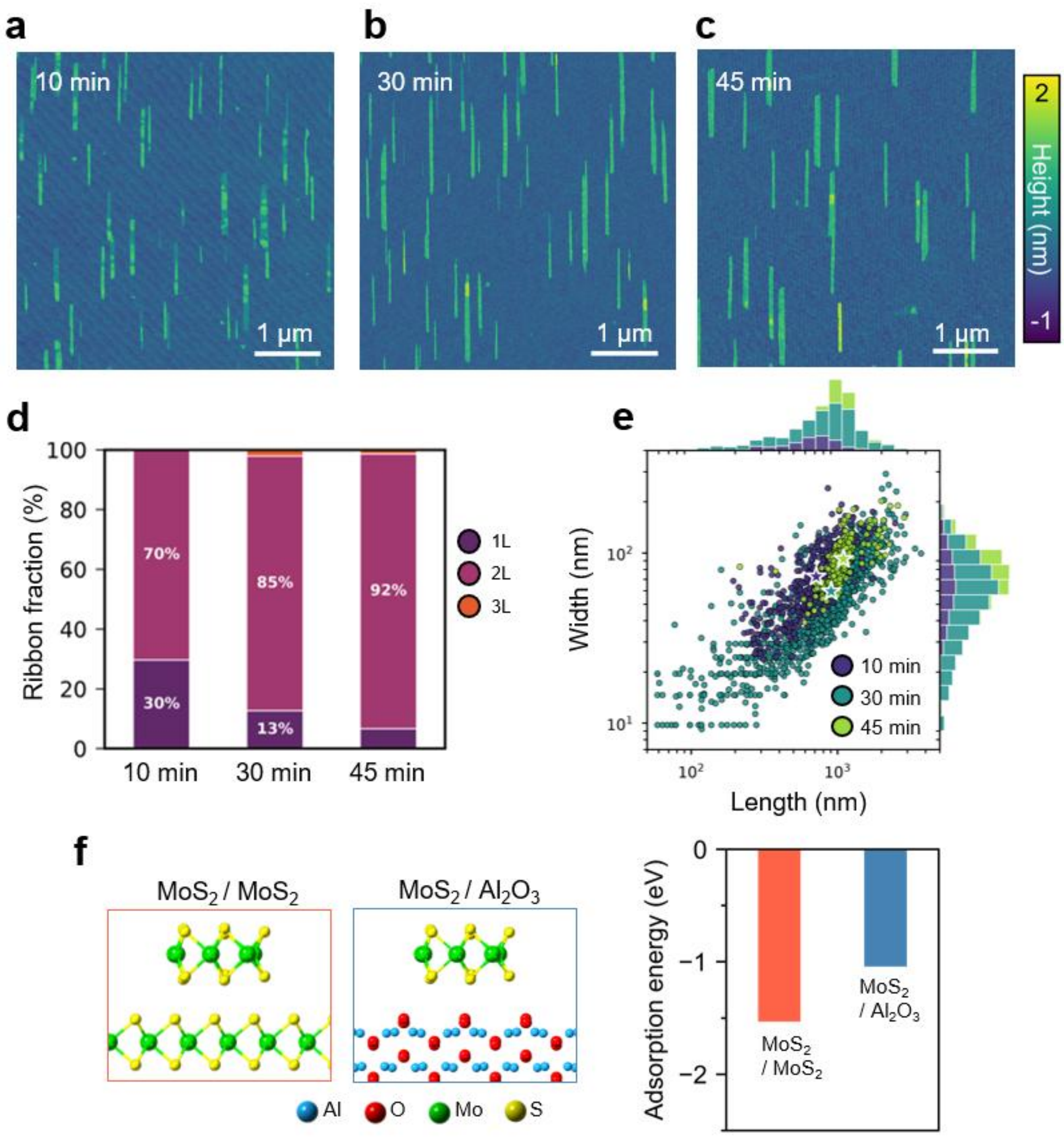

**Fig. 2. Layer-by-layer growth of bilayer $MoS_2$ NRs. a–c** AFM images of $MoS_2$ NR samples prepared with different periods of heating $MoO_3$ feedstock at 640 ºC. **d** Evolution of the number of layers in the aligned $MoS_2$ NRs while increasing the $MoO_3$ heating time. **e** Distributions of NR width and length. Stars indicate the average values for each layer number. **f** Atomic models used for the DFT calculation and the calculated adsorption energies.

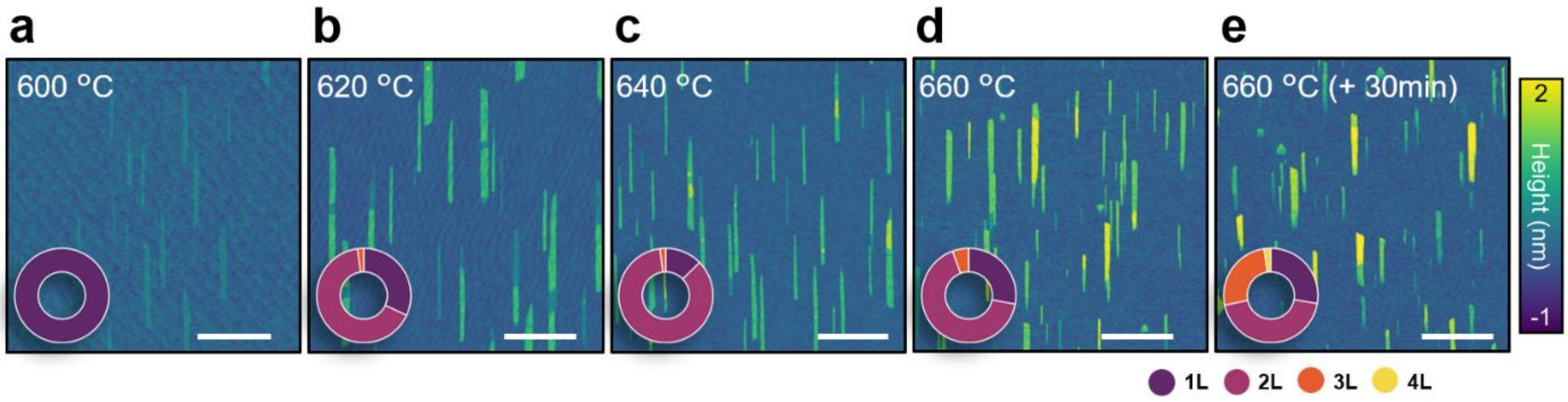

**Fig. 3. Increasing NR layer number by controlling $MoO_3$ supply. a–e** AFM images of NRs grown under different $MoO_3$ temperatures. Sample shown in **a** was obtained at 600 °C for 10 min. **b-d** $MoO_3$ was heated at 600 °C for 10 min, followed by increasing the temperature to 620, 640, and 660 °C, respectively, and maintained for an additional 30 min. **e** $MoO_3$ temperature was kept at 660 °C for additional 60 min. All scale bars are 1 μm.

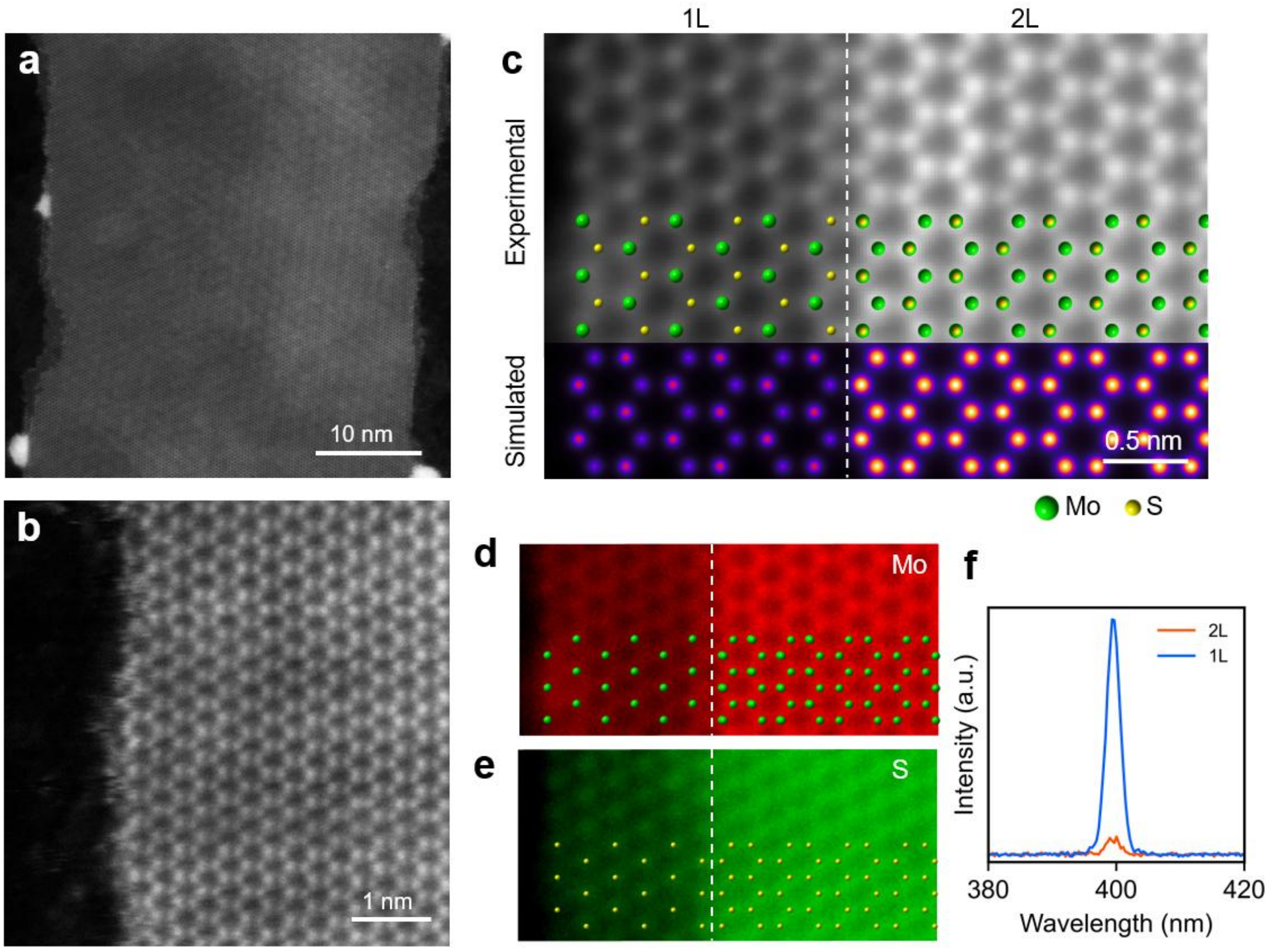

**Fig. 4. Atomic structure and stacking order of a bilayer $MoS_2$ NR.** Low-magnification (**a**) and atomic-resolution (**b**) STEM images of a single-crystalline bilayer $MoS_2$ NR. **c** Atomic-resolution STEM image and corresponding simulated image of the side edge of the bilayer NR, where the termination of the upper layer exposes a monolayer $MoS_2$ region. Corresponding EELS elemental maps of Mo (**d**) and S (**e**). **f** SHG spectra of representative monolayer and bilayer NRs.

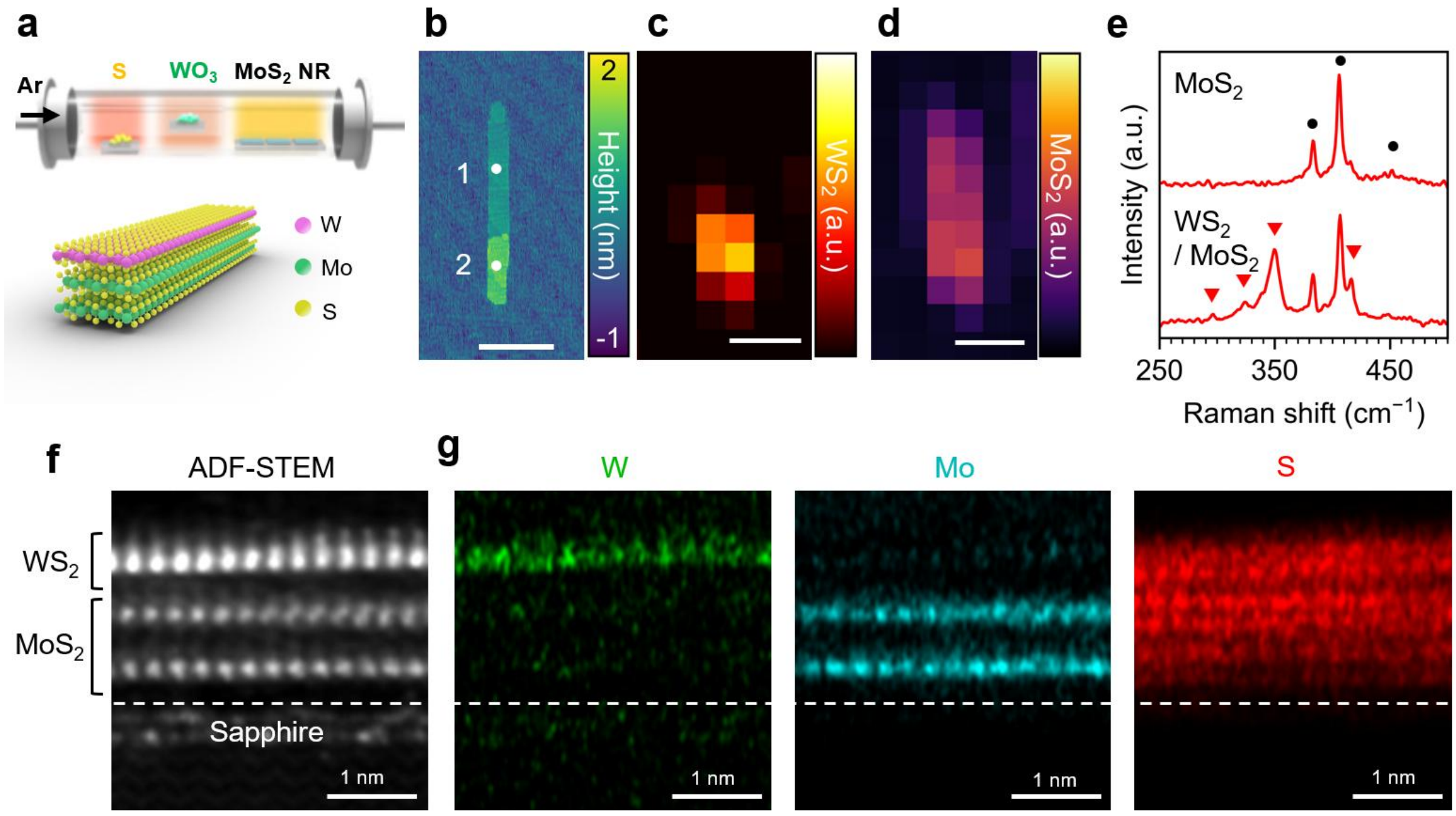

**Fig. 5. Synthesis of hetero-NRs.** **a** CVD setup (upper panel) and schematic of a hetero-NR with $WS_2$ grown on bilayer $MoS_2$ NR (lower panel). AFM image (**b**) and the corresponding Raman intensity mappings of $MoS_2$ (**c**) and $WS_2$ (**d**). Scale bars are 500 nm. **e** Raman spectra of points 1 and 2 marked in **b**. Filled circles and triangles indicate the peaks from $MoS_2$ and $WS_2$, respectively. **f** ADF-STEM image of the cross-section of a hetero-NR. **g** EELS mapping images of W, Mo, and S, confirming the successful synthesis of the $WS_2/MoS_2$ heterostructure.

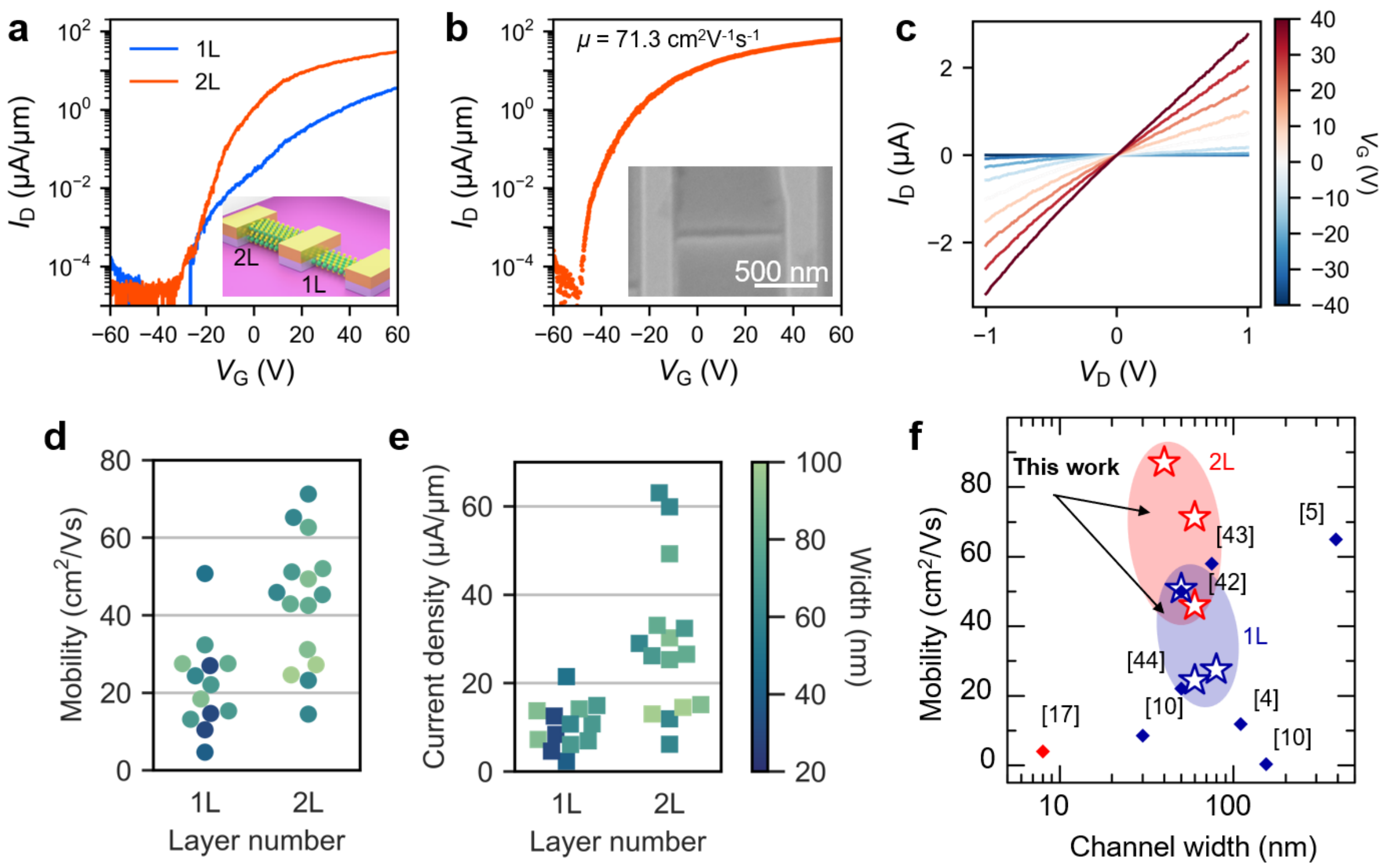

**Fig. 6. Superior device performance of bilayer NRs compared with monolayers. a** Transfer characteristic of FETs with monolayer and bilayer channels in the same NR. Inset shows a schematic of the device structure. Transfer (**b**) and output (**c**) curves of a FET with a bilayer NR channel. Inset in **b** is a scanning electron microscope (SEM) image of the measured device. Distributions of carrier mobilities (**d**) and current densities (**e**) collected from monolayer and bilayer NR FETs. **f** Benchmark of carrier mobilities and channel widths of NR-FETs. Stars show the present results, and symbols in and red and blue express the monolayer[4,5,10,42–44] and bilayer[17]results, respectively.

# Supplementary Information

## Layer-by-layer growth of highly aligned $MoS_2$ nanoribbon arrays

Kaichi Yamamoto[1], Pablo Solís-Fernández[2,*], Haiming Sun[3], Yanlin Gao[4,5], Yeri Lee[6], Yushan Tseng[7], Aika Uchida[2], Masahiro Hara[8], Sunmin Ryu[6], Yung-Chang Lin[3,9], Susumu Okada[4,5], Kazu Suenaga[3], Hiroki Ago[1,2,10,*]

[1] *Interdisciplinary Graduate School of Engineering Science, Kyushu University, Fukuoka 816-8580, Japan*
[2] *Faculty of Engineering Science, Kyushu University, Fukuoka 816-8580, Japan*
[3] *SANKEN (The Institute of Scientific and Industrial Research), The University of Osaka, Osaka 567-0047, Japan*
[4] *Department of Physics, Graduate School of Science and Technology, University of Tsukuba, Tsukuba 305-8577, Japan*
[5] *Tsukuba Institute for Advanced Research (TIAR), University of Tsukuba, Tsukuba 305-8565, Japan*
[6] *Department of Chemistry, Pohang University of Science and Technology (POSTECH), Gyeongbuk 37673, Republic of Korea*
[7] *Graduate School of Science and Technology, Kumamoto University, Kumamoto 860-8555, Japan*
[8] *Faculty of Advanced Science and Technology, Kumamoto University, Kumamoto 860-8555, Japan*
[9] *Research Institute of Core Technology for Materials Innovation, National Institute of Advanced Industrial Science and Technology (AIST), Tsukuba 305-8565, Japan*
[10] *Center for Semiconductors and Device Ecosystem (CSeDE), Kyushu University, Fukuoka 816-8580, Japan*

[*]Corresponding authors: Hiroki Ago (ago.hiroki.974@m.kyushu-u.ac.jp), Pablo Solís-Fernández (solisfernandez.pablo.kyudai@gmail.com)

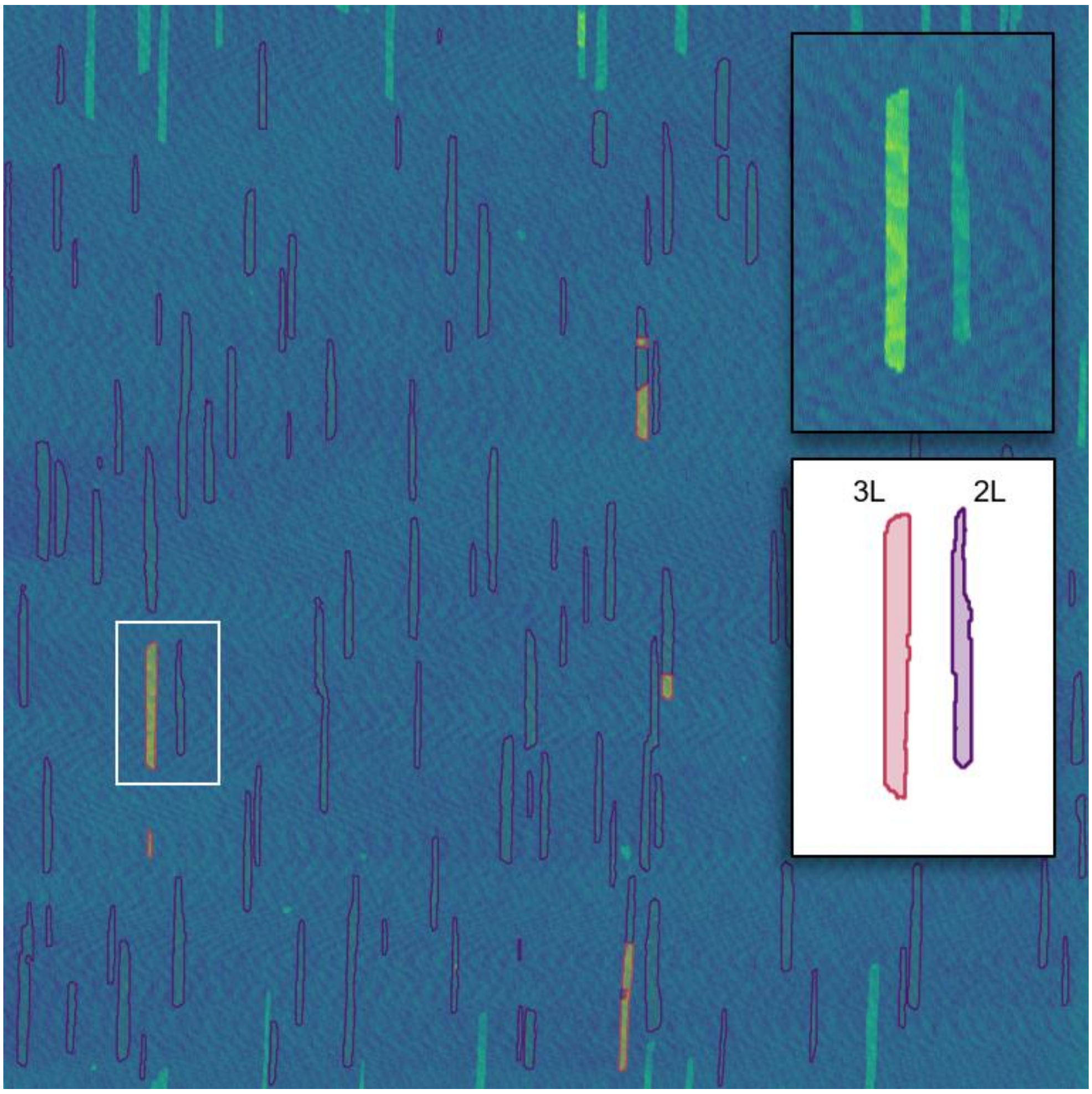

**Supplementary Fig. 1. Segmentation in AFM topographic images.** Each NR in the AFM image is highlighted with a different color according to its layer number: bilayer (purple) and trilayer (red). Note that NRs overlapping the image boundaries were excluded from the analysis because accurate information could not be obtained. The insets show a zoom of the AFM image (top inset) and the NR contours (bottom inset) of the region highlighted by the white rectangle.

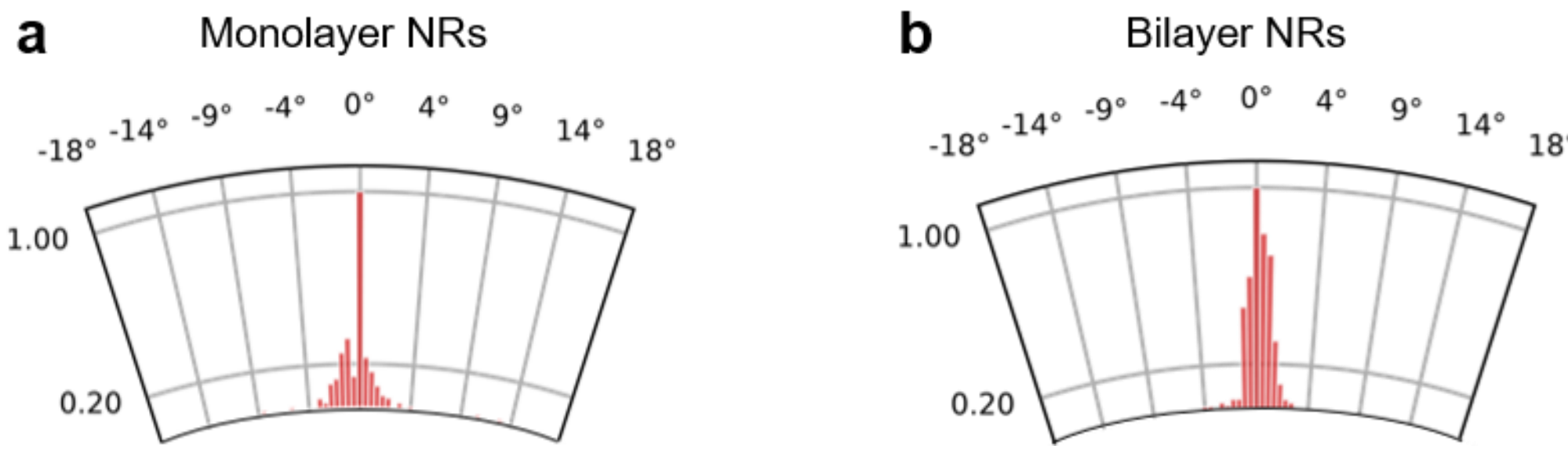

**Supplementary Fig. 2. Orientation distributions of the monolayer and bilayer NRs. a,b** Polar-coordinate histograms of the NR alignment in the AFM images shown in Fig. 1c,d, with the dominant angle set to 0°, evidencing a high degree of alignment. Based on the STEM observations, the dominant alignment direction is assigned to the $[1\bar{1}00]$ direction.

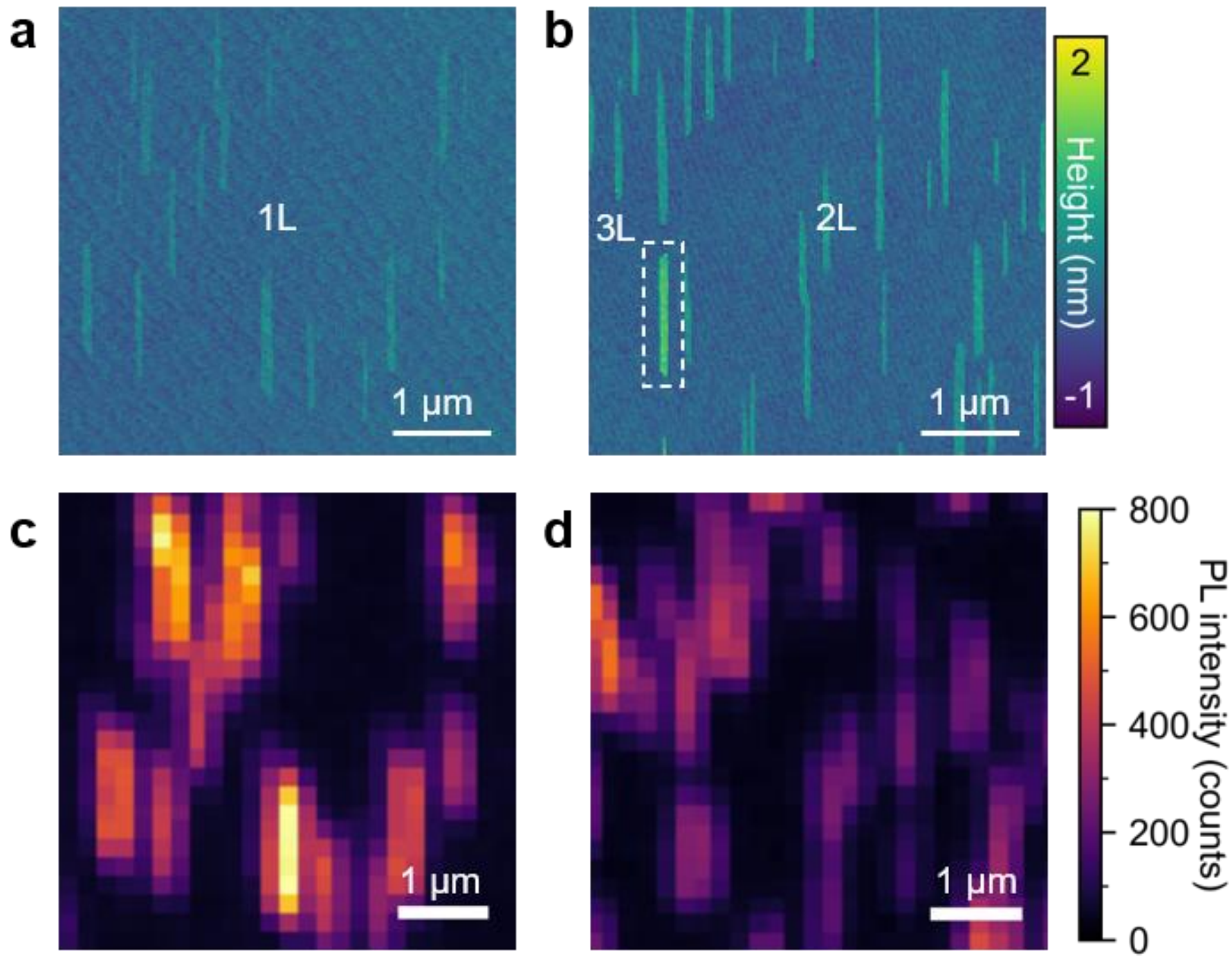

**Supplementary Fig. 3**. **AFM images and PL intensity mappings acquired at the same region.** AFM images of the region corresponding to the Raman mappings shown in Fig. 1j: monolayer (**a**) and bilayer (**b**) NRs. A trilayer NR is highlighted by a white dashed rectangle in **b**. PL intensity mappings of monolayer (**c**) and bilayer (**d**) NRs.

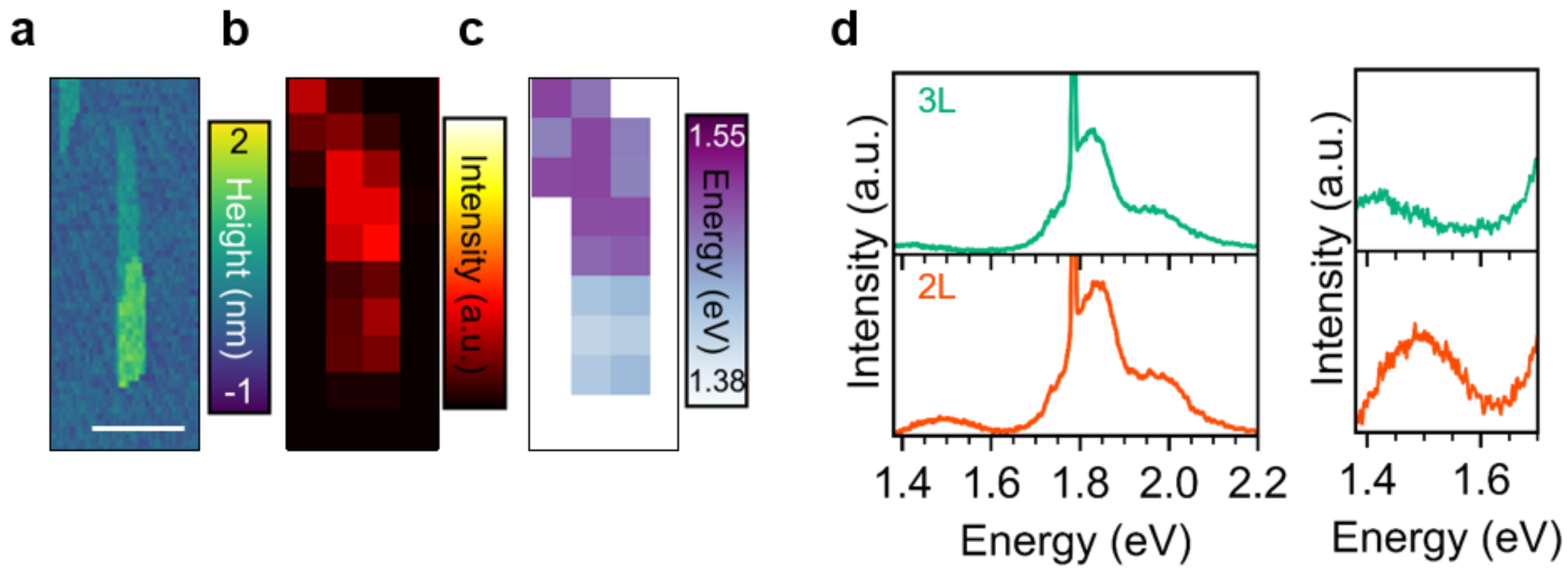

**Supplementary Fig. 4 Comparison of PL spectra of bilayer and trilayer NRs. a** AFM image of a NR containing both bilayer and trilayer regions. Scale bar is 500 nm. Intensity mapping (**b**) and peak-position mapping (**c**) for the indirect-transition PL peak. **d** PL spectra of bilayer (bottom) and trilayer (top) regions. Compared with the bilayer, the trilayer exhibits a weaker indirect-transition PL peak shifted to lower energy.

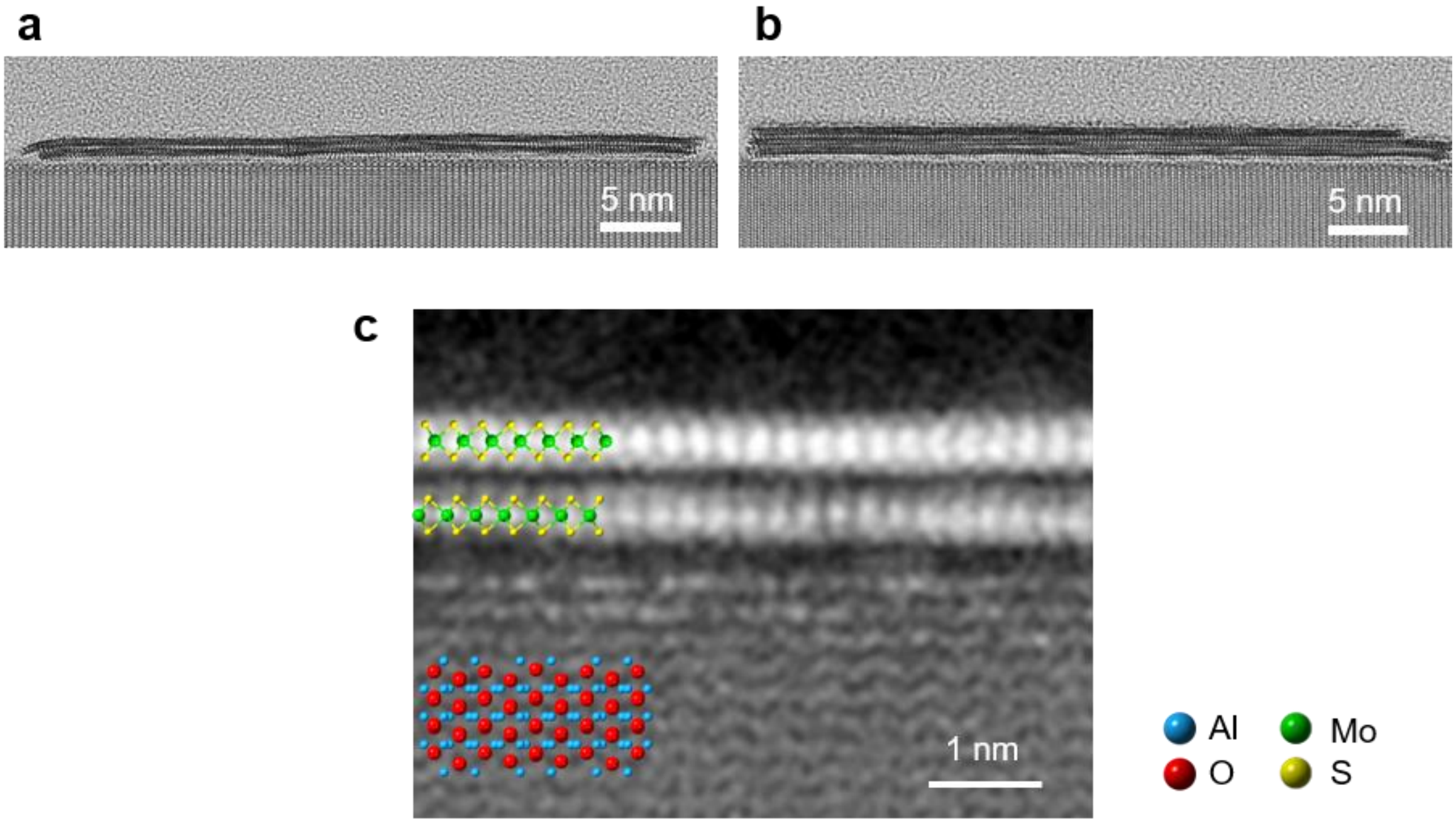

**Supplementary Fig. 5. Cross-section of bilayer and trilayer NRs.** Cross-sectional TEM images of bilayer (**a**) and trilayer (**b**) NRs. **c** STEM image of bilayer NRs on a-plane sapphire. The atomic models of 2H-stacked bilayer NR and sapphire m-plane are overlapped in the image.

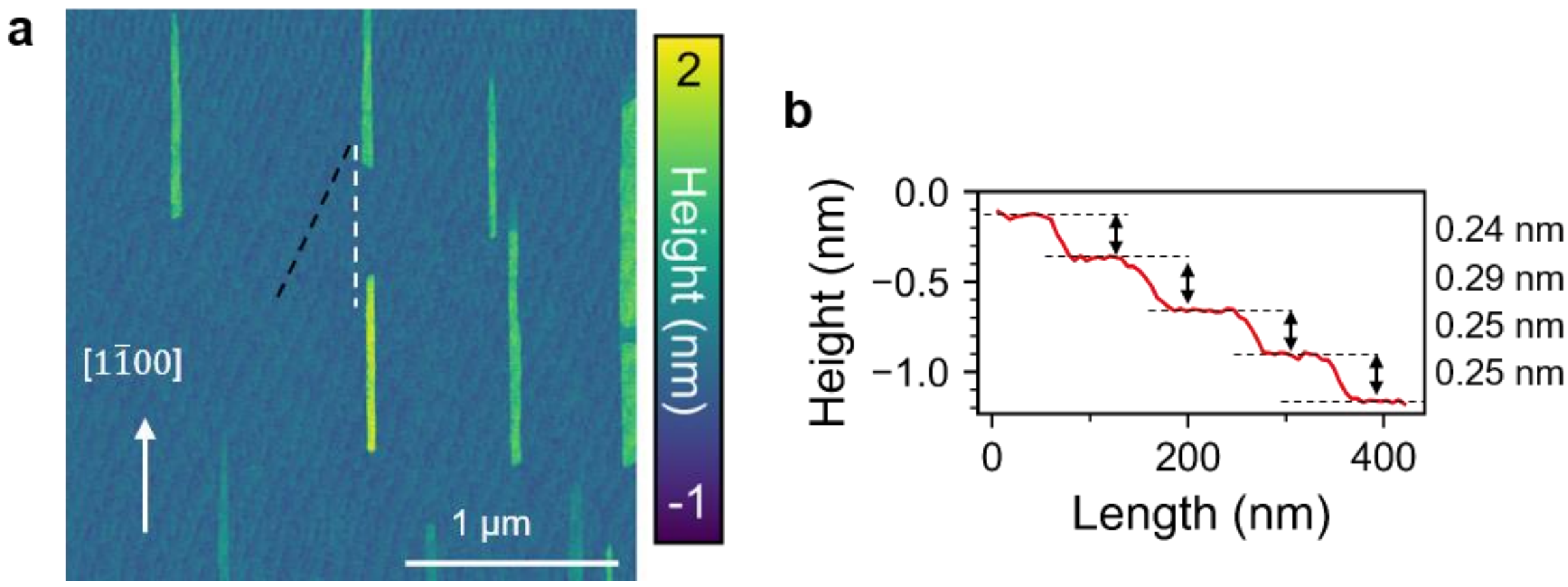

**Supplementary Fig. 6. Negligible influence of surface steps on alignment direction of the $MoS_2$ NRs. a** AFM image of as-grown NRs on a-plane sapphire. Misalignment between the NRs (white dashed line) and the steps (black dashed line) is clearly recognized. **b** Line profile perpendicular to the steps. The height of the sapphire steps is ~0.25 nm.

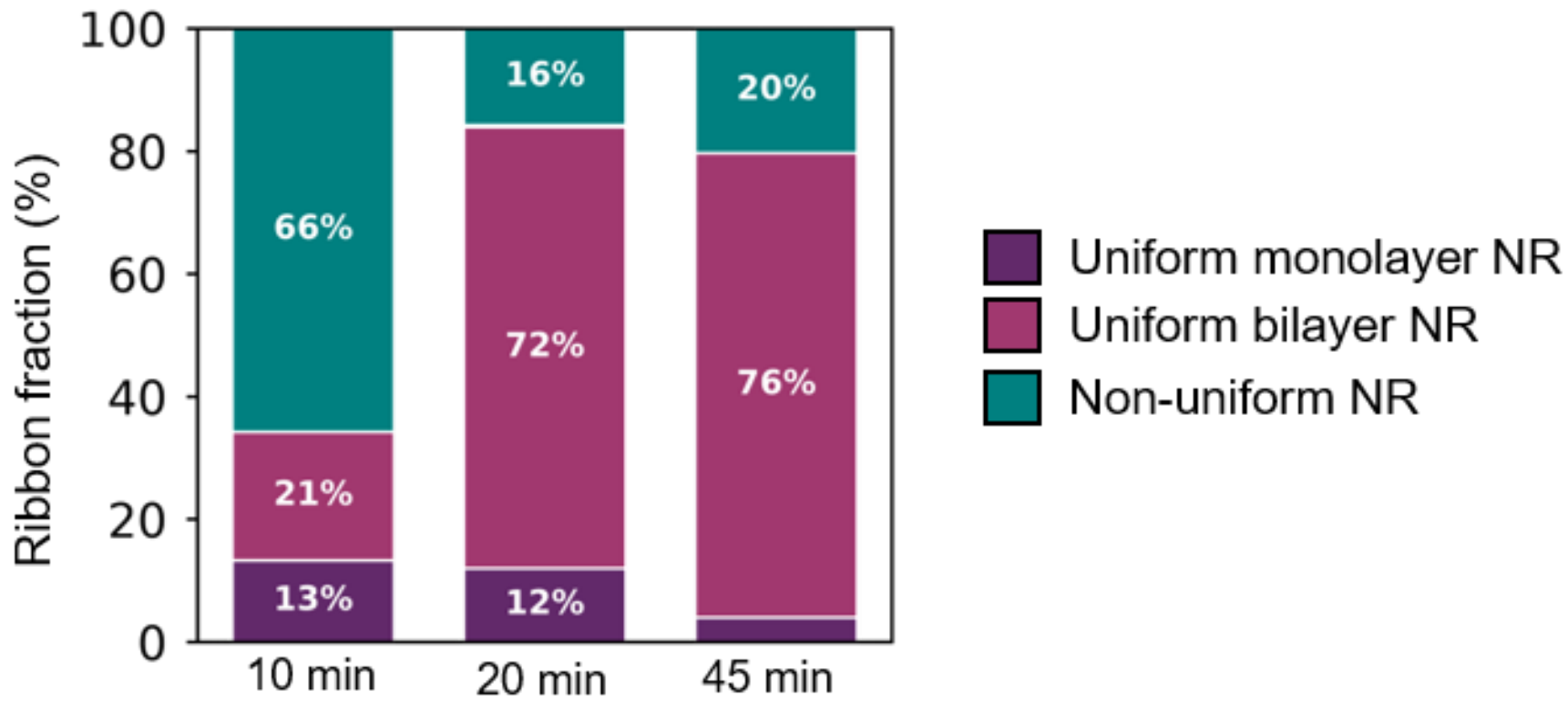

**Supplementary Fig. 7. Distribution of uniform NRs with different layer numbers grown for different durations at $MoO_3$ temperature of 640 °C.** NRs in which the bilayer (monolayer) domain occupies most of the NR area are classified as uniform bilayer (monolayer) NRs. NRs that do not meet this criterion were classified as non-uniform NRs. As the growth time increases, the second-layer coverage increases, leading to a higher fraction of uniform bilayer NRs and a lower fraction of uniform monolayer and non-uniform NRs. This trend provides evidence of the layer-by-layer growth.

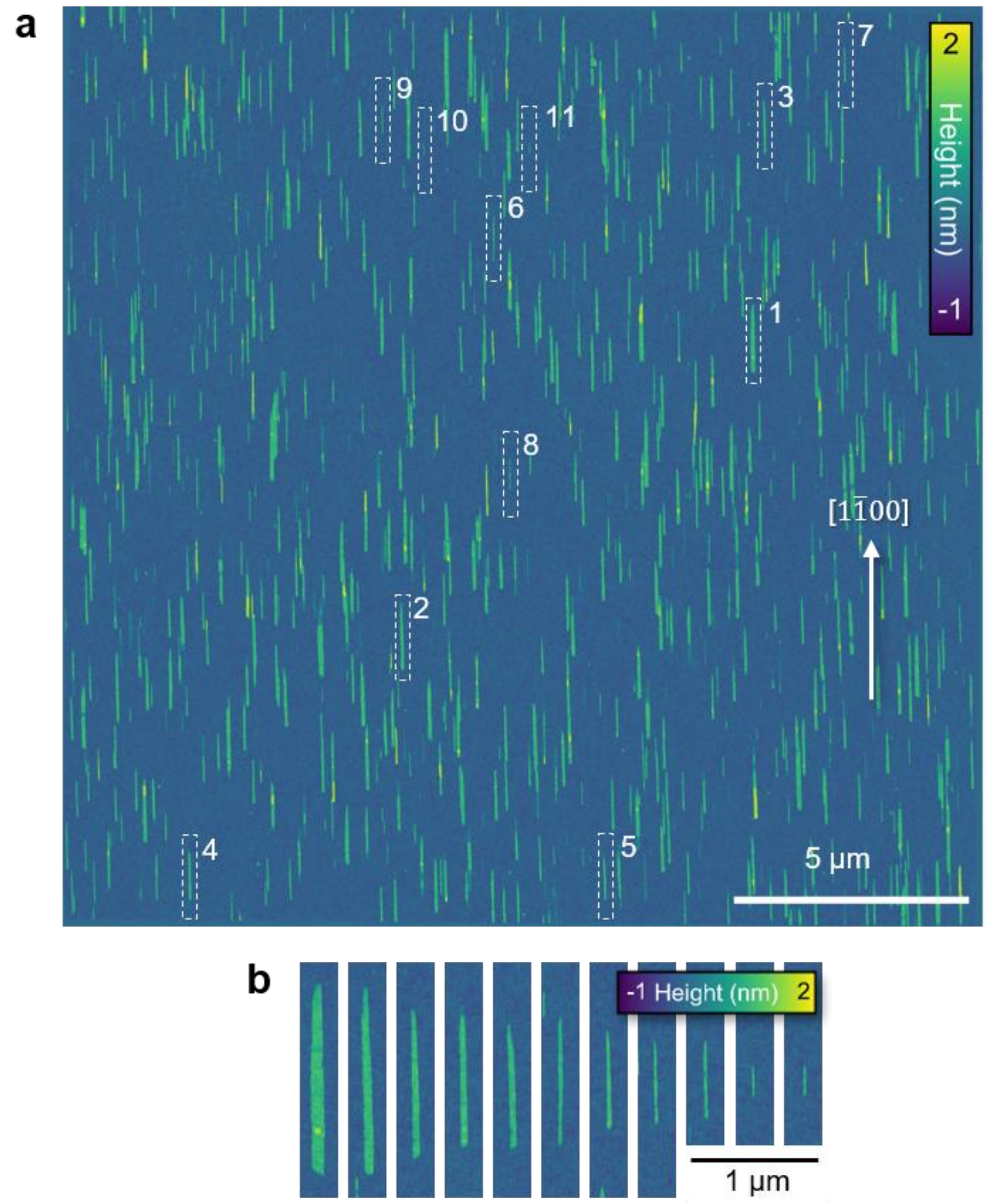

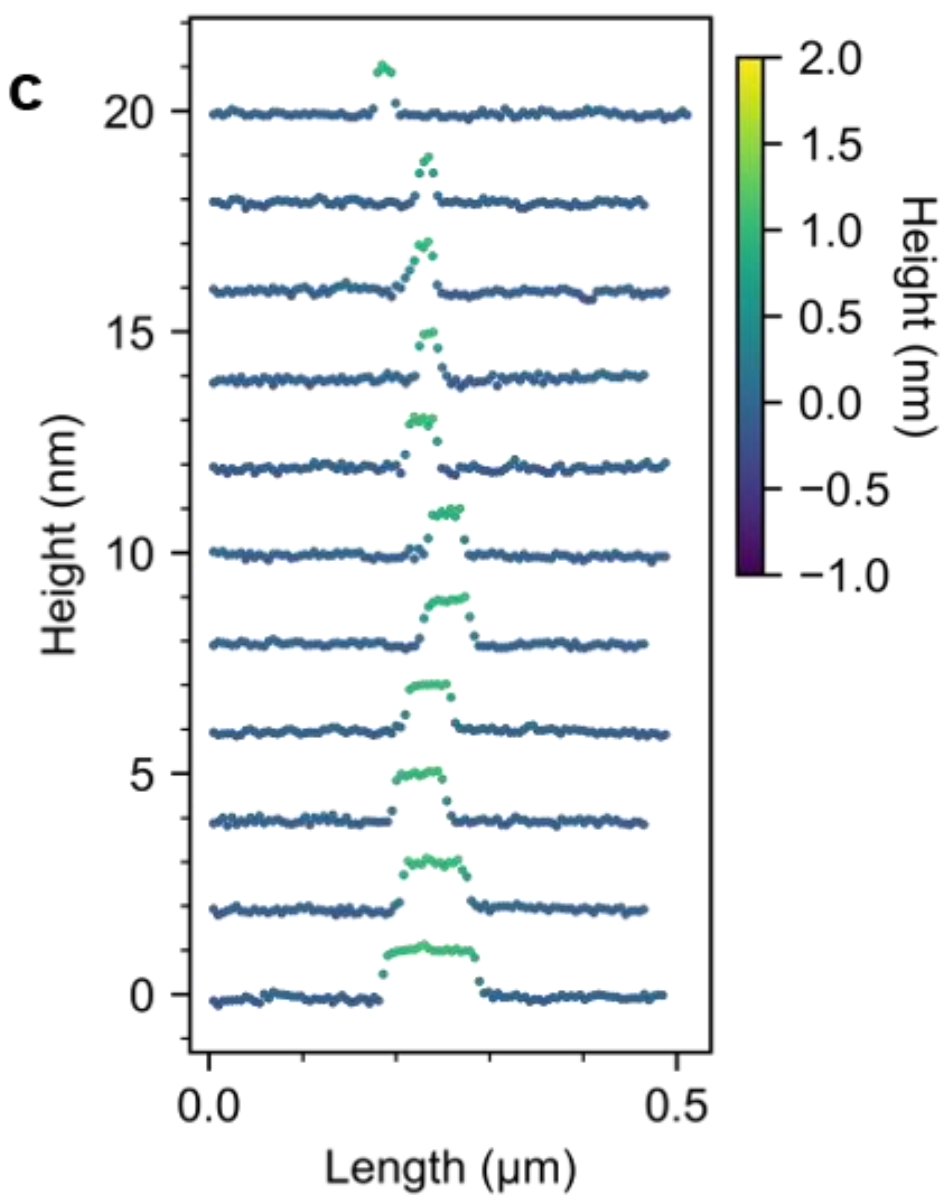

**d**

| Ribbon | Width (nm) | Height (nm) |
|---|---|---|
| #1 | 98.0 | 1.1 |
| #2 | 72.6 | 1.1 |
| #3 | 58.7 | 1.1 |
| #4 | 48.8 | 1.0 |
| #5 | 53.8 | 1.0 |
| #6 | 39.0 | 1.0 |
| #7 | 34.1 | 1.1 |
| #8 | 24.5 | 1.1 |
| #9 | 29.3 | 1.1 |
| #10 | 19.6 | 1.0 |
| #11 | 19.5 | 1.0 |

**Supplementary Fig. 8. Size of the aligned bilayer $MoS_2$ NRs. a** Large-area AFM image of the bilayer NRs. **b** AFM images of isolated bilayer NRs. Height profiles (**c**) and measured size (**d**) of each NR highlighted in **a**. The profiles in **c** are vertically offset for clarity, and the color represents the actual height.

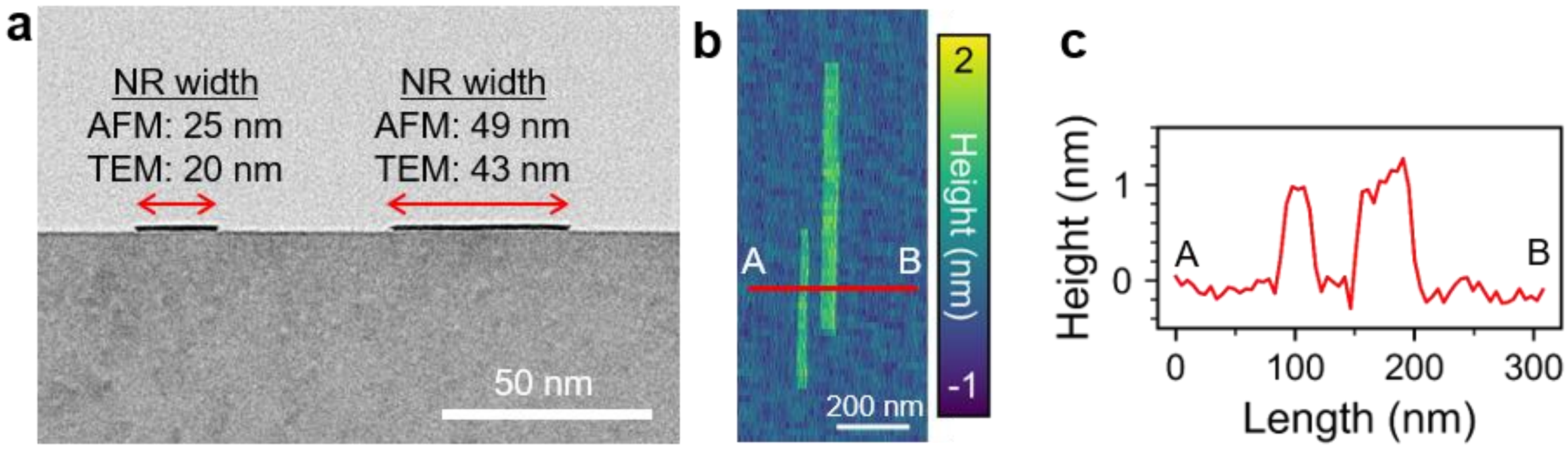

**Supplementary Fig. 9**. **Effect of tip convolution in AFM images. a** Cross-sectional TEM image of two bilayer NRs. **b** The corresponding top-view AFM image. **c** Height profile along line A–B in **b**. As seen in **a**, the NR widths measured from the TEM image are 20 and 43 nm for the left and right NRs, respectively, while the AFM image gives 25 and 49 nm widths. This result indicates that the NR widths measured by AFM are overestimated due to the effect of tip convolution.

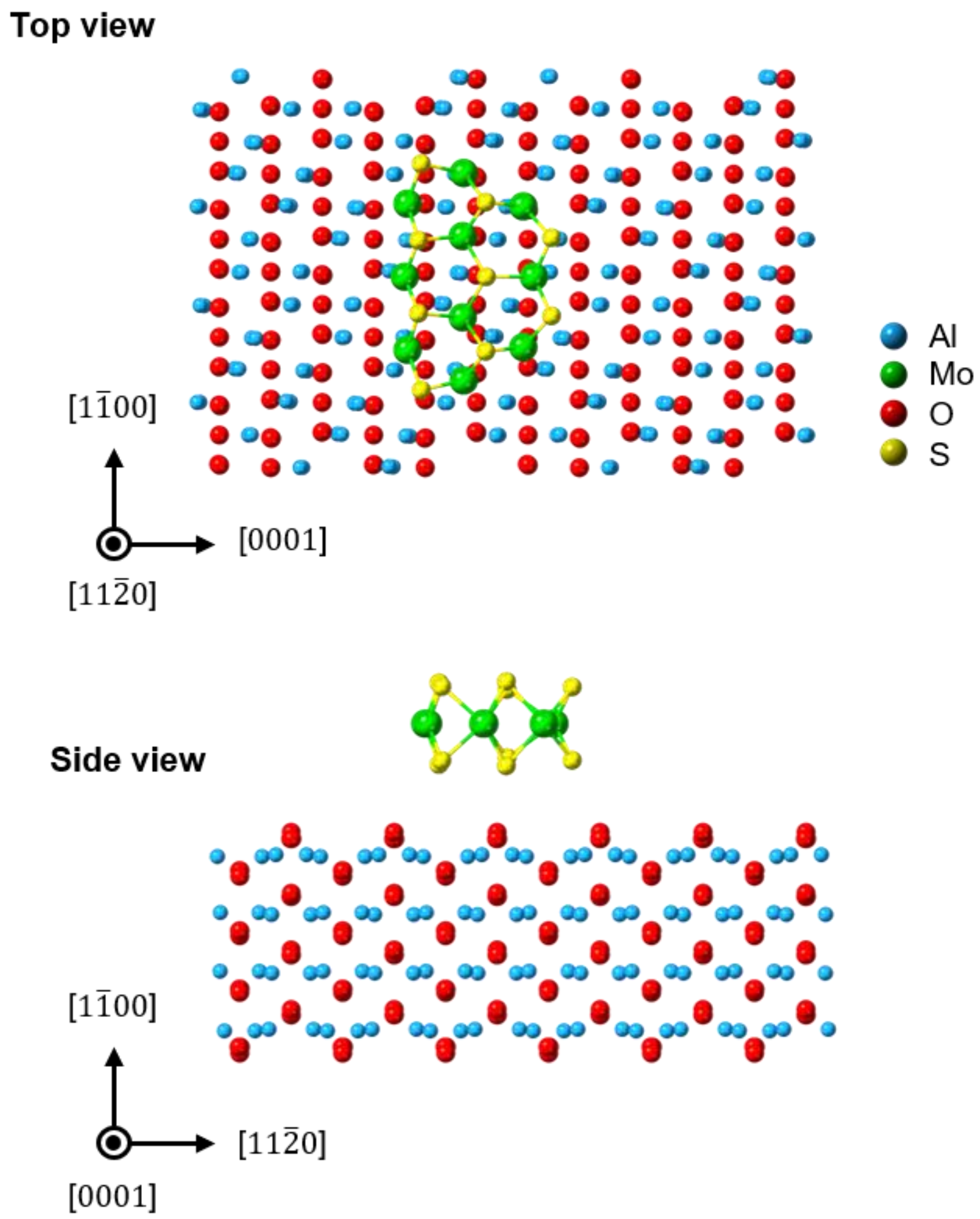

**Supplementary Fig. 10.** Atomic models used for the DFT calculations of the adsorption energy of a $MoS_2$ flake on a-plane sapphire.

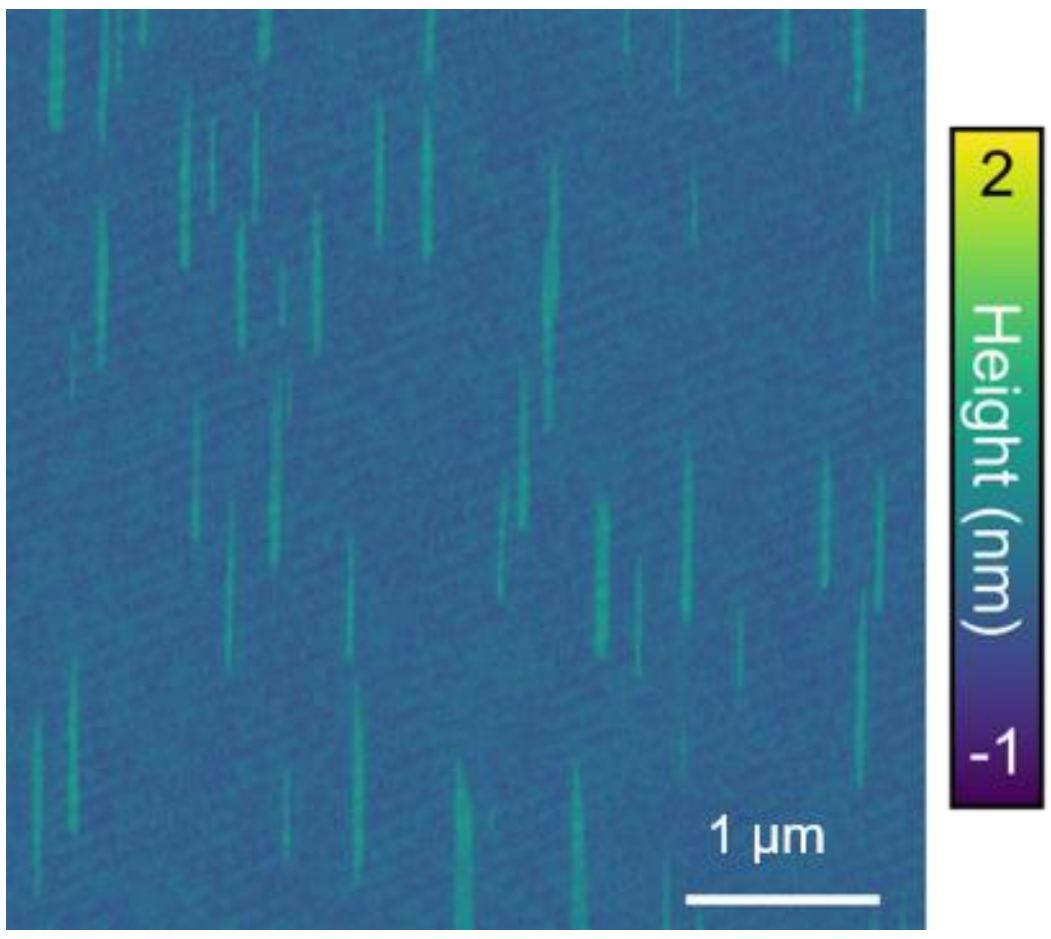

**Supplementary Fig. 11.** AFM image of monolayer NRs grown while keeping the $MoO_3$ temperature at 600 °C for 20 min. Here, the growth time was doubled compared with that shown in Fig. 1c, but the NRs still remain monolayer.

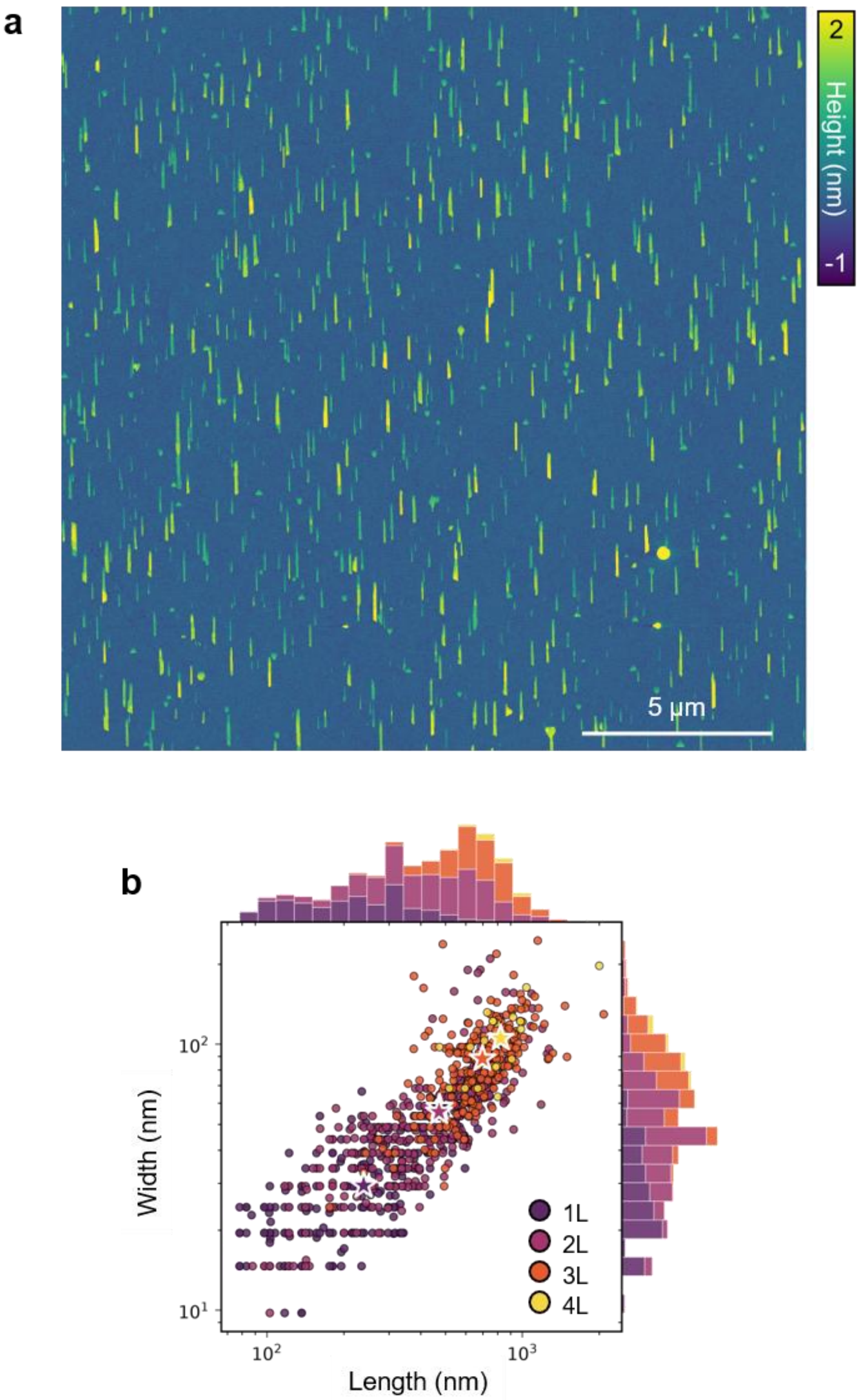

**Supplementary Fig. 12**. AFM image (**a**) and distribution (**b**) of width and length of NRs grown at the $MoO_3$ temperature of 660 °C for 75 min, yielding high fractions of trilayer and four-layer NRs. Stars in **b** indicate the average values for each layer number.

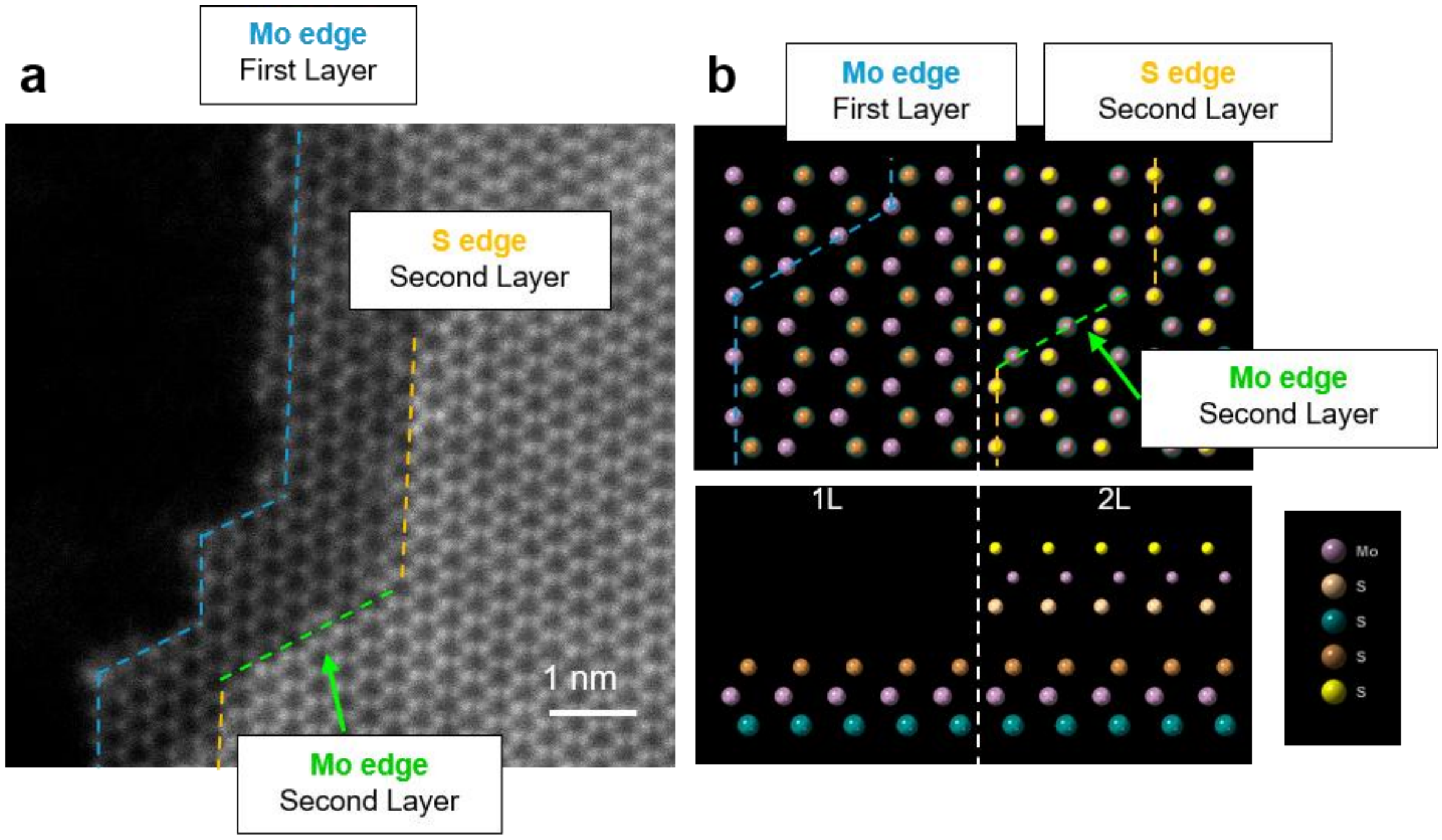

**Supplementary Fig. 13**. **Reconstruction at the S-terminated edge of the second layer. a** STEM image of an NR edge. **b** Atomic models of $MoS_2$ with 2H stacking. The second-layer edge is globally parallel to the ZZ direction of the $MoS_2$, whereas some local edge segments exhibit ZZ-Mo edge (green dashed lines) in addition to the ZZ-S edge expected for the second-layer edge in the 2H stacking configuration (orange dashed lines).

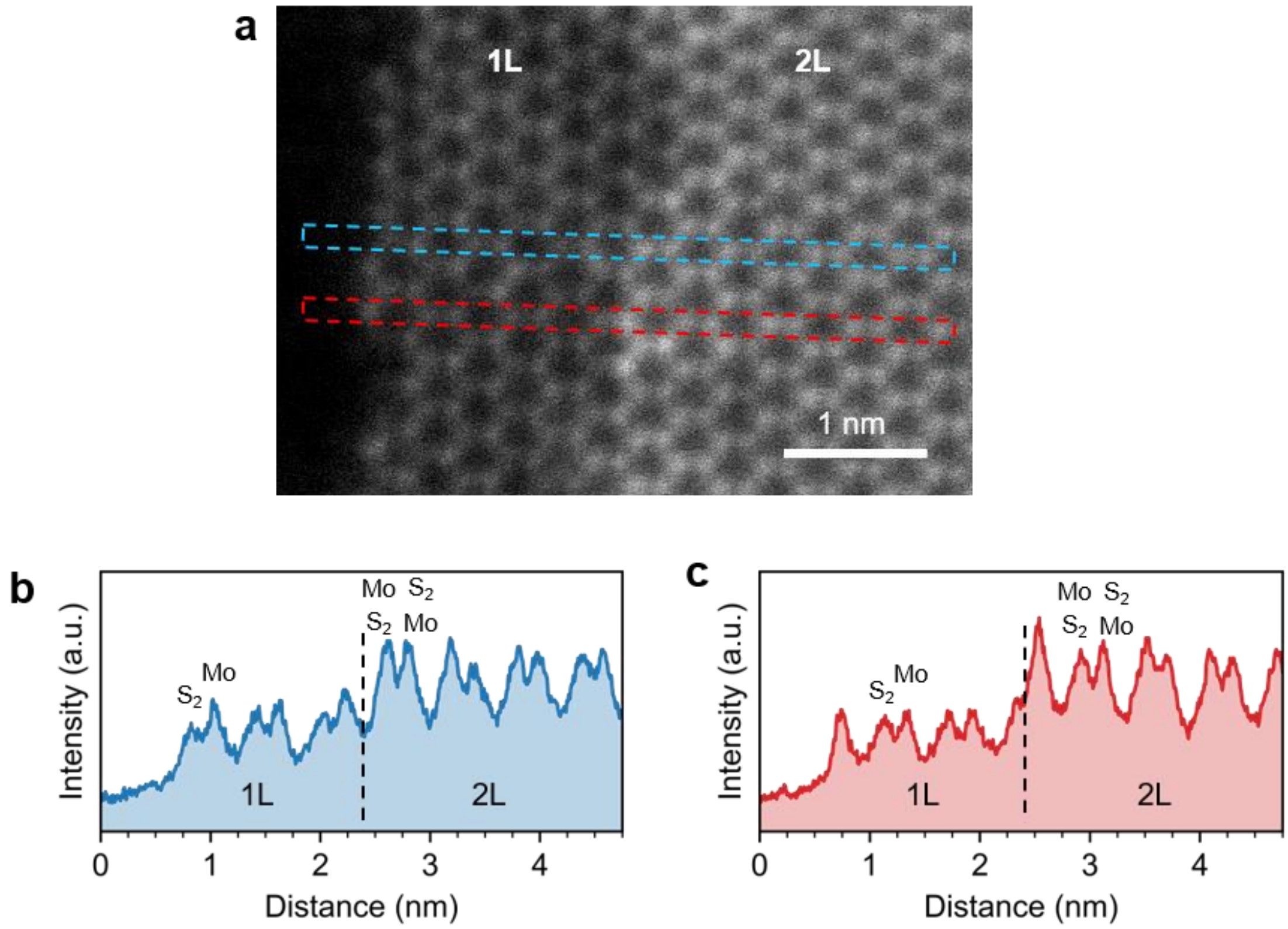

**Supplementary Fig. 14. a** STEM image of a NR edge. Intensity profiles along the blue (**b**) and red (**c**) dashed rectangles shown in **a**. The intensity increases from monolayer to bilayer, and the intensity of each atomic column in the bilayer region is consistent with Mo+$S_2$, confirming the 2H stacking configuration.

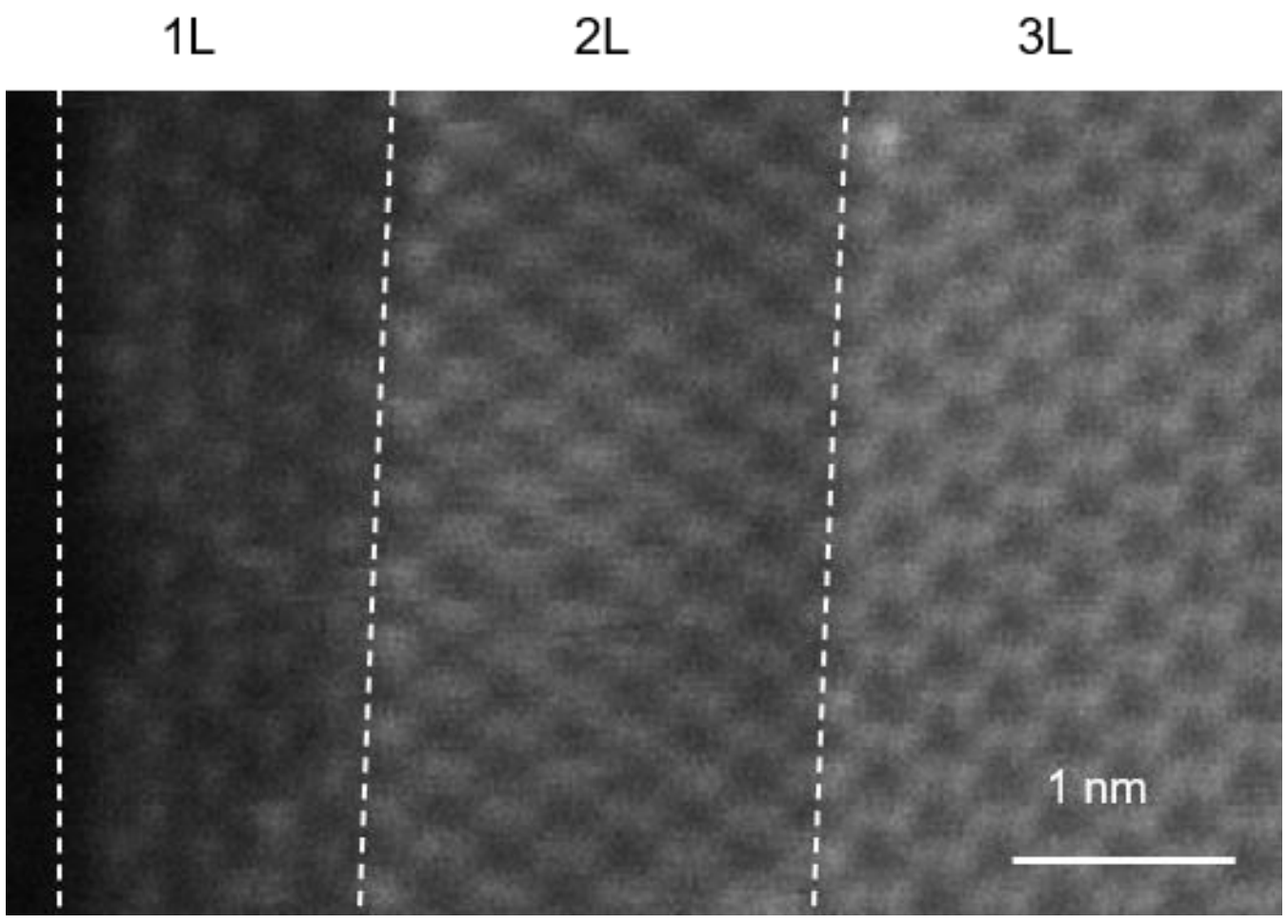

**Supplementary Fig. 15**. STEM image of a NR edge showing the first, second, and third layers in a region where the edges do not coincide, indicating that the trilayer also possesses 2H stacking.

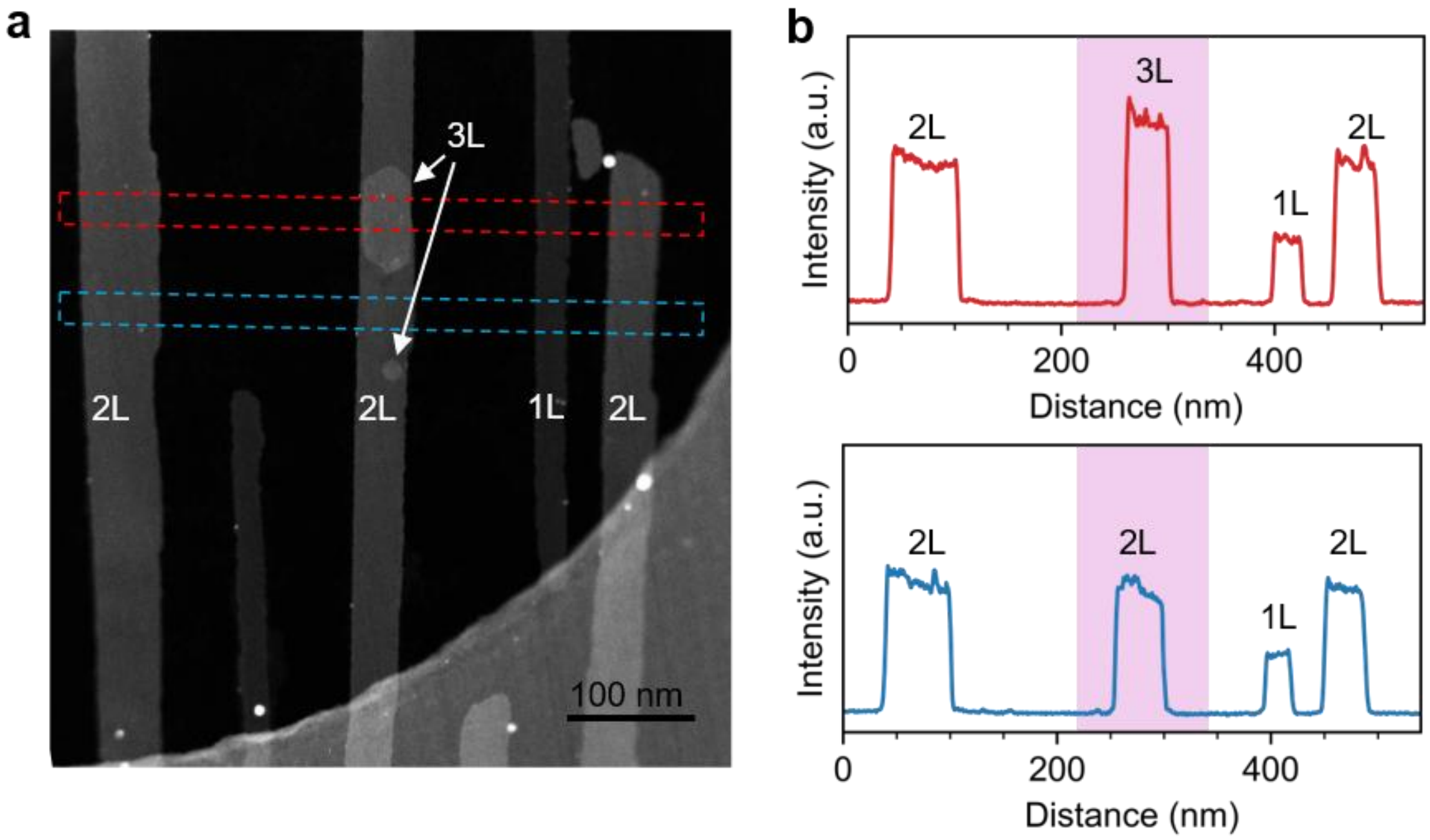

**Supplementary Fig. 16**. **a** Low-magnification STEM image of the region that includes monolayer and bilayer NRs. The third-layer islands also formed on the second layer, as indicated by white arrows. **b** Intensity profiles collected along the dashed rectangles in **a**, confirming the third layer is formed on the second layer.

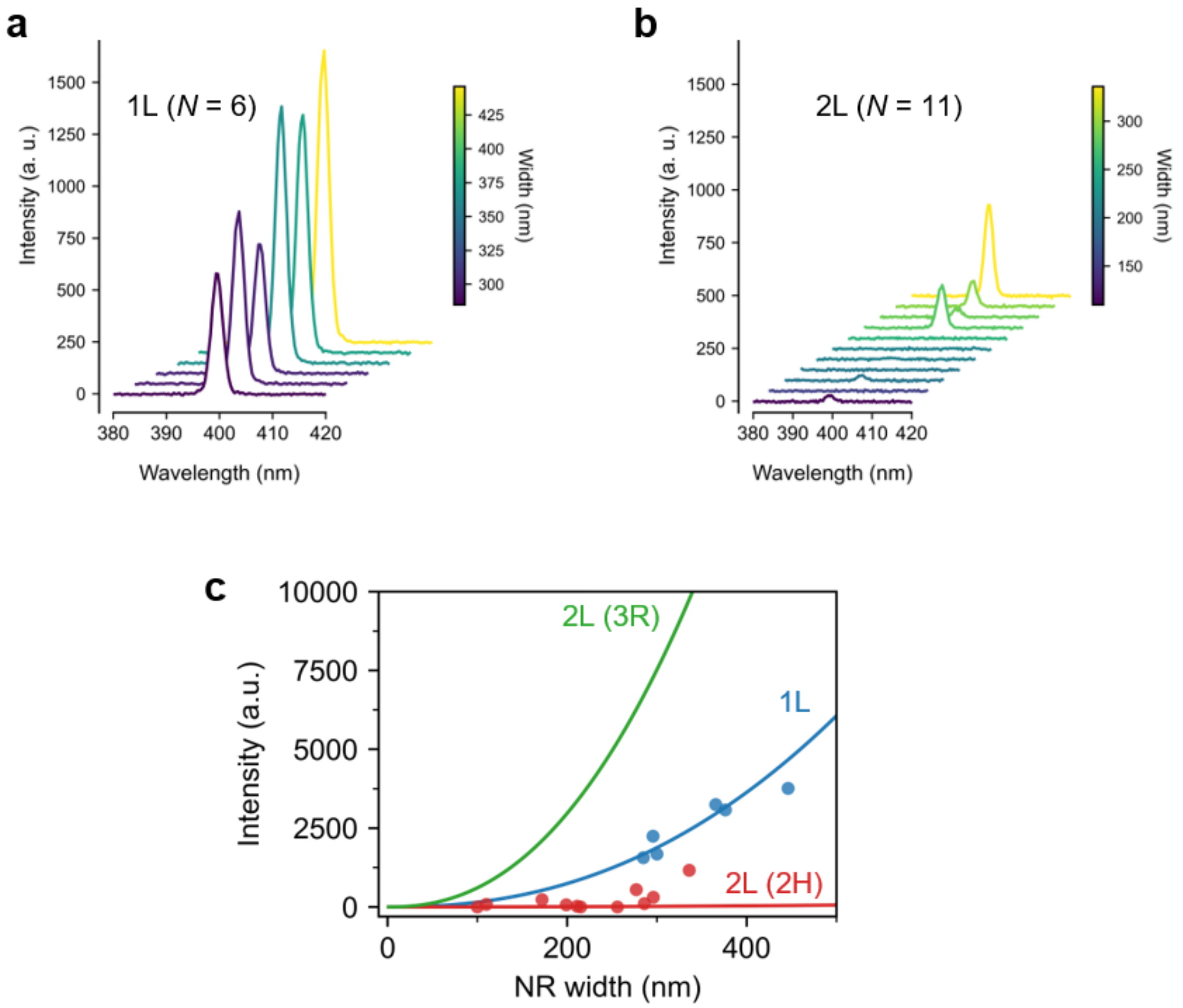

**Supplementary Fig. 17**. **Stacking characterization by SHG.** SHG spectra obtained from monolayer (**a**) and bilayer (**b**) NRs. Each spectrum was collected from different NRs. The profiles are vertically and horizontally offset for clarity. **c** Dependence of NR width on SHG intensity for monolayer (blue closed circles) and bilayer (red closed circles). The SHG intensities for bilayer NRs with 2H (red solid line) and 3R (green solid line) stacking are estimated from the fitted SHG intensity of monolayer NRs (blue solid line). The SHG intensity of the bilayer NRs is not as strong as that expected for 3R stacking and does not depend on the NR width, supporting that the obtained bilayer NRs have a 2H stacking configuration.

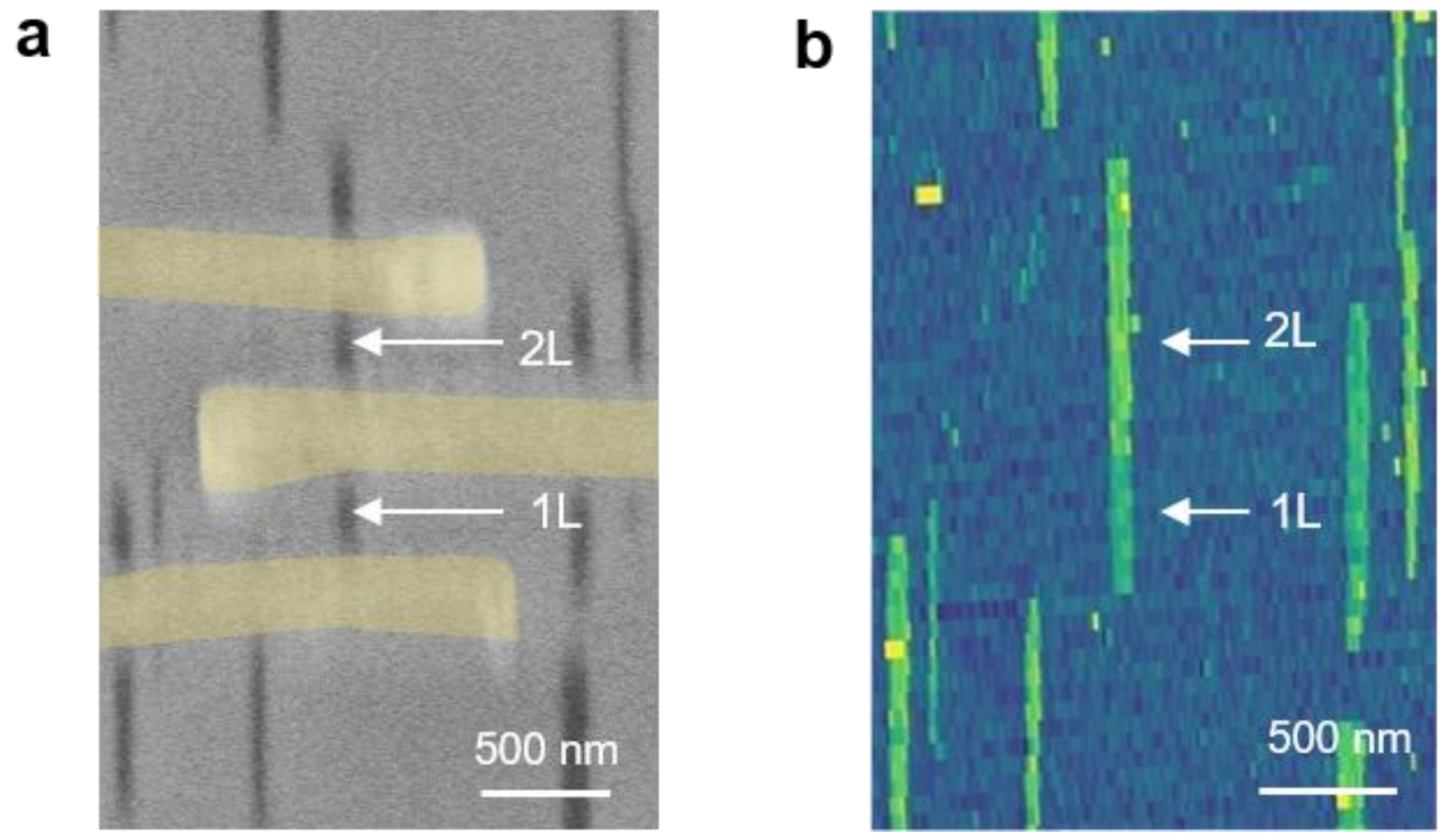

**Supplementary Fig. 18**. **a** SEM image of the FET containing both 1L and 2L regions. This device data is presented in Fig. 6a. **b** Corresponding AFM image of the NR measured before depositing the electrodes, proving the coexistence of 1L and 2L regions.

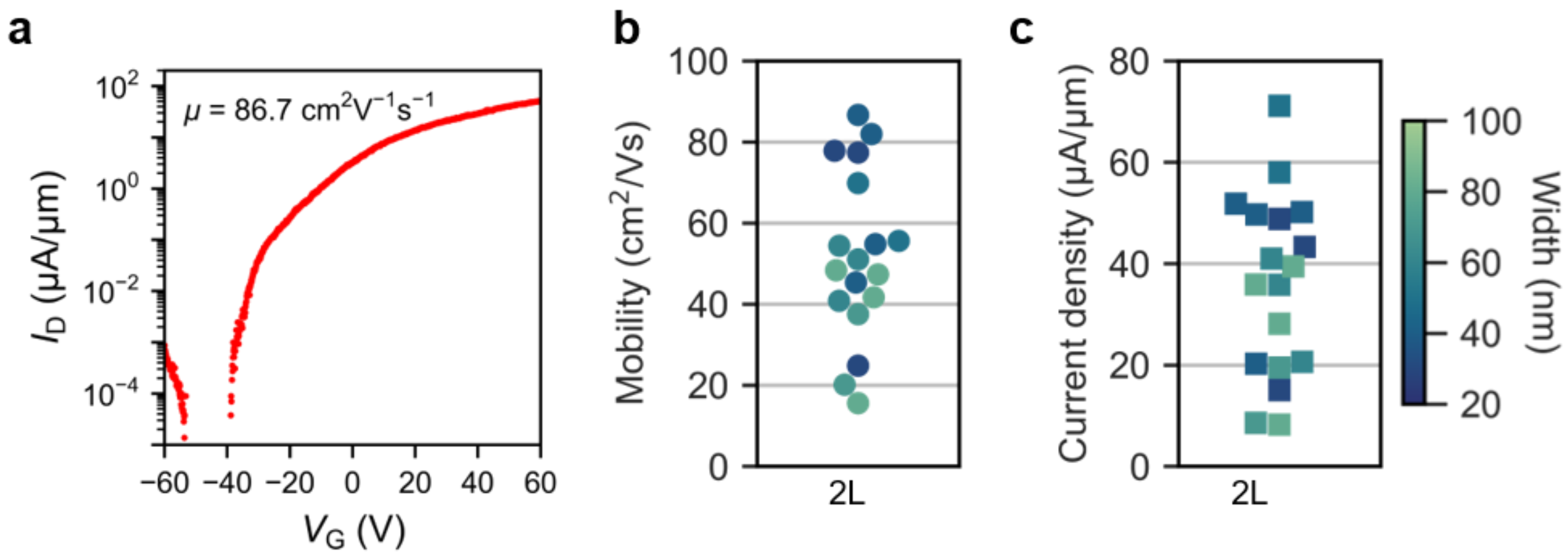

**Supplementary Fig. 19**. **Device performance of FETs with Sb contact. a** Transfer characteristic of an FET with a bilayer NR channel with a width and a length of 40 nm and 900 nm, respectively. Au/Sb electrodes were employed. Mobilities (**b**) and current densities (**c**) collected from the Sb-contact bilayer NR FETs.

**Supplementary Table 1**. Comparison of device performance of TMD NRs

| Ref. | Layer number | $W_{ch}$ (nm) | $L_{ch}$ (nm) | Mobility ($cm^2V^{-1}s^{-1}$) | Current density (μA/um) | $V_D$ (V) | Gate dielectric | Fabrication method |
|---|---|---|---|---|---|---|---|---|
| [17] | 2 | 8 | 400 | 4 | 9 | 0.1 | $SiO_2$ (285 nm) | CVD |
| [5] | 1 | 390 | 1000 | 65 | 10 | 0.5 | $HfO_2$ (15 nm) + 1L-h-BN | CVD |
| [4] | 1 | 110 | 700 | 11.8 | 0.91 | 0.5 | $SiO_2$ (300 nm) | CVD |
| [12] | 1 | 8 | 1000 | | 22 | 1 | $SiO_2$ | CVD |
| [10] | 1 | 30 | 160 | 8.53 | | 1 | $SiO_2$ (285 nm) | Scanning probe lithography + $XeF_2$ etching + $Al_2O_3$ passivation |
| [10] | 1 | 154 | 1420 | 0.36 | | 1 | $SiO_2$ (285 nm) | Scanning probe lithography + $XeF_2$ etching |
| [42] | 1 | 50 | 500 | 50 | 38 | 1 | $SiO_2$ (300 nm) | Eelectron beam lithography + $SF_6$ etching |
| [43] | 1 | 43 | 300 | | 620 | 1 | $SiO_2$ (96 nm) | Electron beam lithography + $XeF_2$ etching |
| [43] | 1 | 75 | 300 | 58 | 400 | 1 | $SiO_2$ (96 nm) | Electron beam lithography + $XeF_2$ etching |
| [44] | 1 | 50 | 500 | 22 | 1 | 0.1 | $SiO_2$ (300 nm) | Eelectron beam lithography + $SF_6$ etching |
| This work | 2 | 60 | 900 | 71.3 | 63.1 | 1 | $SiO_2$ (300 nm) | CVD |
| This work | 2 | 60 | 900 | 45.9 | 28.9 | 1 | $SiO_2$ (300 nm) | CVD |
| This work | 2 | 40 | 900 | 87 | 52 | 1 | $SiO_2$ (300 nm) | CVD |